\documentclass{aa}  
\usepackage{graphicx}
\usepackage[squaren,Gray]{SIunits}
\usepackage{footnote}
\usepackage{txfonts}
\usepackage{xcolor}
\usepackage{enumitem}
\usepackage{enumerate}
\usepackage{hyperref}
\usepackage{wasysym}
\usepackage{pifont}
\usepackage[referable]{threeparttablex}
\usepackage{diagbox}
\usepackage{multirow}
\usepackage{subeqnarray}
\usepackage{cases}
\usepackage{lscape}

\renewlist{tablenotes}{enumerate}{1}
\makeatletter
\setlist[tablenotes]{label=\tnote{\alph*},ref=\alph*,itemsep=\z@,topsep=\z@skip,partopsep=\z@skip,parsep=\z@,itemindent=\z@,labelindent=\tabcolsep,labelsep=.2em,leftmargin=*,align=left,before={\footnotesize}}
\makeatother

\definecolor{db}{HTML}{2A2181}
\definecolor{dv}{HTML}{227310}
\definecolor{dr}{HTML}{8D1717}
\definecolor{or}{HTML}{784B12}
\definecolor{vin}{HTML}{731013}
\definecolor{gris}{HTML}{7E7684}
\definecolor{spring}{HTML}{00830B}
\definecolor{jade}{HTML}{39AD7E}
\definecolor{jv}{HTML}{BFBF00}
\definecolor{mauve}{HTML}{773535}
\definecolor{darkorange}{HTML}{854611}
\definecolor{bleuvert}{HTML}{2E9C65}
\newcommand{\uc}{u_\mathrm{c}}
\newcommand{\lc}{l_\mathrm{c}}

\newcommand{\bu}{{\bf u}}

\newcommand{\bF}{{\bf F}}
\newcommand{\bB}{{\bf B}}
\newcommand{\bb}{{\bf b}}
\newcommand{\bn}{{\boldsymbol\nabla}}
\newcommand{\bd}{{\boldsymbol\nabla^2}}
\newcommand{\Le}{\mathrm{Le}}
\newcommand{\Ek}{\mathrm{Ek}}
\newcommand{\Em}{\mathrm{Em}}
\newcommand{\KE}{\mathrm{KE}}
\newcommand{\ME}{\mathrm{ME}}
\newcommand{\Ro}{\mathrm{Ro}}
\newcommand{\Roh}{\mathrm{Ro}_\mathrm{t}}
\newcommand{\st}{\sigma_\mathrm{t}}
\newcommand{\Oo}{\Omega_\mathrm{o}}
\newcommand{\e}[1]{\cdot10^{#1}}
\newcommand{\h}{\mathrm{e}}
\newcommand{\sm}{\sigma_\mathrm{max}}
\newcommand{\scm}{\hat\sigma_\mathrm{max}}

\renewcommand{\theenumi}{\roman{enumi}}
\renewcommand{\fm}{f_\mathrm{mag}}
\renewcommand{\fh}{f_\mathrm{hydro}}
\newcommand{\bfo}{}
\newcommand{\bfc}{}

\begin{document} 
   \title{Does magnetic field impact tidal dynamics inside the convective zone of low-mass stars along their evolution?}
   \subtitle{}
   \author{A. Astoul\inst{1}\fnmsep\thanks{aurelie.astoul@cea.fr}
    \and
     S. Mathis\inst{1}
	\and
	C. Baruteau\inst{2}
    \and
    F. Gallet\inst{3}
    \and
    A. Strugarek\inst{1}
    \and
    K. C. Augustson\inst{1}
    \and
    A. S. Brun\inst{1}
    \and
    E. Bolmont\inst{4}
          }
   \institute{AIM, CEA, CNRS, Universit\'e Paris-Saclay, Universit\'e Paris Diderot, Sorbonne Paris Cit\'e, F-91191 Gif-sur-Yvette, France
       \and
	IRAP, Observatoire Midi-Pyr\'en\'ees, Universit\'e de Toulouse, 14 avenue Edouard Belin, 31400 Toulouse, France
    \and 
	Univ. Grenoble Alpes, CNRS, IPAG, 38000 Grenoble, France
    \and
	Observatoire de Gen\`eve, Universit\'e de Gen\`eve, 51 Chemin des Maillettes, CH-1290 Sauverny, Switzerland 
\\
}
   \date{Received XXXX; accepted YYYY}
 
  \abstract
   {The dissipation of the kinetic energy of wave-like tidal flows within the convective envelope of low-mass stars is one of the key physical mechanisms that shapes the orbital and rotational dynamics of short-period exoplanetary systems. Although low-mass stars are magnetically active objects, the question of how the star's magnetic field impacts large-scale tidal flows and the excitation, propagation and dissipation of tidal waves still remains open.
}
   {Our goal is to investigate the impact of stellar magnetism on the forcing of tidal waves, and their propagation and dissipation in the convective envelope of low-mass stars as they evolve.}
   {We have estimated the amplitude of the magnetic contribution  to the forcing and dissipation  of tidally induced magneto-inertial waves throughout the structural and rotational evolution of low-mass stars (from M to F-type). For this purpose, we have used detailed grids of rotating stellar models computed with the stellar evolution code STAREVOL. The amplitude of dynamo-generated magnetic fields is estimated via physical scaling laws at the base and the top of the convective envelope.}
   {We find that the large-scale magnetic field of the star has little influence on the excitation of tidal waves in the case of  nearly-circular orbits and coplanar hot-Jupiter planetary systems, but that it has a major impact on \bfo the way waves are dissipated\bfc. Our results therefore indicate that a full magneto-hydrodynamical treatment of the propagation and dissipation of tidal waves is needed to properly assess the impact of star-planet tidal interactions  throughout the evolutionary history of low-mass stars hosting short-period massive planets.}
   {}
   \maketitle
%
\section{Introduction}
Over the last two decades, a large variety of exoplanetary systems has been discovered, primarily through photometric  transit and radial velocity observations  \citep[e.g.][]{P2018}. 
Among these systems, several populations of exoplanets orbit very close to their host stars (with orbital periods of  a few days), such as  hot-Jupiters, super-Earths, and mini-Neptunes  \citep{MQ1997,S2014}. 
In these short period systems as well as in tight binary stars, \bfo tides induced by each other's body \bfc drive  the rotational and orbital evolutions of the system through dissipation mechanisms \citep[the so-called tidal dissipation, see e.g.][]{Z1977,H1980,ZB1989,GN1989,WS2002}. 

Usually, tides are split  into two components  following the work of  \cite{Z1966a,Z1966b,Z1966c,Z1975,Z1977}. First,  the equilibrium tide is the quasi-hydrostatic response of the main body  to tidal perturbations induced by the companion. It is materialised by a large-scale deformation in the perturbed body, that is displayed as a near-equatorial tidal bulge in the direction of the companion. In addition, waves are excited  in its interior as the equilibrium tide  is not an exact solution to the equations of motion, providing an additional driving force  \citep{O2014}. The tidally-forced waves correspond to the  dynamical tides.
The convective zone (CZ)  of a perturbed rotating body is the seat of inertial waves that get dissipated by turbulent friction \citep[e.g.][]{OL2004,OL2007}, while the radiative zone supports gravito-inertial waves which are dissipated by thermal damping and turbulent friction \citep[e.g.][]{Z1975, TP1998,GD1998,BO2010}.  

The dissipation of the dynamical tides in the CZ is very efficient in young stars (pre-main sequence and early main sequence) whereas for older stars the dissipation of the equilibrium tide dominates mainly because of the slower stellar rotation, as demonstrated for instance by \cite{BM2016} \citep[see also ][]{SB2017,G2017,BM2018}.  
The efficiency of the tidal excitation and viscous dissipation of inertial and gravito-inertial waves can be measured with the tidal quality factor $Q$.  This quantity reflects the  fact that the perturbed body undergoes a forced oscillation and dissipates a fraction of the associated energy during each oscillation period. 
It has been evaluated by \citet{OL2004,OL2007}, with and without  differential rotation \cite[see also][for this particular topic]{G2016a,G2016b}, in the case of giant planets and solar-type stars. 
Recently, \cite{M2015} and \cite{G2017} explored the influence of mass, age and rotation based on the frequency-averaged dissipation  estimates of \cite{O2013} to understand the behaviour of tidal dissipation along the evolution of stars. They emphasised that the variation of these parameters could  drastically modify the strength of tidal dissipation with a higher frequency-averaged tidal dissipation for low-mass stars. Similarly, \cite{BG2017} highlighted the importance of the stellar metallicity.  
Within this context, the variations of tidal dissipation along stellar evolution have a strong impact on the orbital architecture  of compact planetary systems and the planet survival  \citep{BM2016,BR2019}.

In the aforementioned studies of star-planet tidal interactions, an important ingredient is  missing though: stellar magnetism. In the Sun and solar-like stars, magnetism is revealed  by external magnetic features such as sunspots, prominences or flares  \citep{DL2009}. The magnetic fields  of solar-like stars originates from a powerful dynamo mechanism, sustained  by turbulent convection and differential rotation in the convective  envelope of the star \citep{BB2017}.
Recent endeavours have been carried out to assess the effects  of magnetism on tidally-excited inertial waves in stars \citep{W2016,W2018,LO2018}. 
In the presence of a magnetic field, tidal waves excited in the CZ become magneto-inertial waves. Moreover, they feel the magnetic tension of the large-scale magnetic field which affects their propagation and dissipation \citep{F2008}. 
These magnetically modified inertial waves have a broader  range of propagation frequencies, compared to the hydrodynamical case, and can be dissipated through both viscous and Ohmic processes. Specifically,  the transition between hydrodynamical and magneto-hydrodynamical (MHD) regimes have been explored  in a shearing  box model with a uniform magnetic field \citep{W2016} and in spherical geometry  with both a uniform \bfo field directed along the z-axis \bfc and a dipolar magnetic field \citep{LO2018}. The authors of these studies both stressed that the Lehnert number $\Le$  \citep{L1954} determines how important magnetism is to tidal dissipation.  This dimensionless number compares  the Alfvén velocity to  the rotation speed of the body. 
Additionally, how magnetism influences the effective tidal force to excite magneto-inertial waves remains to be addressed, as mentioned in \citet[Appendix B]{LO2018}.
In practice, given the equation of motion tidal waves are excited by an effective body force  driven by the Coriolis acceleration of   the equilibrium tidal flow in the non-magnetized case  \citep{O2005}. In the presence of a magnetic field, the Lorentz force acting on the equilibrium tide is likely to play a role in the excitation of tidal waves. Thus, this motivates the study of the impact of stellar magnetism on both dissipation and excitation of dynamical tides inside the CZ and along the evolution of low-mass stars. 

The paper is organised as follows. In Section \ref{sec21} we work out  the contribution of magnetism to  tidal forcing and derive  a criterion to assess its importance relative to non-magnetized forces. This criterion features the Lehnert number and thus the magnetic field and the rotation speed of the body, as well as the tidal forcing frequency.  Simple scaling laws are applied in Sect. \ref{sec22} to estimate the dynamo-driven magnetic field in the convective envelope of low-mass stars. Thanks to the stellar evolution code STAREVOL \citep[Sect. \ref{sec23}, see also][and references therein]{AP2019}, the strength of a mean magnetic field is given, in Sect. (\ref{sec24}),  at the base and top of the CZ for various low-mass stars. Thus, we  evaluate in Sect. \ref{sec25} the Lehnert number as a function of age, mass, initial rotation, and radius in the CZ of these stars. 
 We then estimate in Sect. \ref{sec26} the Lehnert number along with the rotation and tidal frequencies in several observed short-period exoplanetary systems to assess the importance of the star's magnetic field on the tidal forcing. Based on our estimates of the Lehnert number, we compare  in Sect. \ref{sec3} the relative  importance of Ohmic over  viscous dissipations of tidally induced magneto-inertial waves throughout the evolution of low-mass stars. In Sect. \ref{sec4}, we examine how small-scale magnetic fields impact tidal forcing. Finally, we present in Sect. \ref{sec_conclu} the conclusions and perspectives of this work.

\section{The Lorentz force influence on tidal forcing}
The purpose of this section  is to quantify the contribution of the stellar magnetic field to  the tidal excitation of magneto-inertial waves in the CZ of low-mass stars.
\subsection{A criterion to settle the importance of magnetism}
\label{sec21}
 We aim for an approach as general as possible, yet we restrict our model adopting a solid-body rotation with an angular frequency $\Omega$. In particular, we do not assume  a specific geometry for the magnetic field. We linearise  the momentum and induction equations to derive  the magnitude of the magnetic tidal forcing (the effective tidal force arising from the Lorentz force) and  compare it with the classical hydrodynamical tidal forcing. We introduce the self-gravitational potential $\Phi_0$, as well as the gravitational potential perturbation  $\Phi$, and the external tidal potential $\Psi$ \citep[see e.g.][]{Z1966a,O2013}. The continuity and entropy equations are as given in Zahn's paper, with the addition of Ohmic heating to the entropy equation.  
The momentum equation for tidal  perturbations in the co-rotating frame can be written as:
\begin{equation}
\begin{aligned}
\rho_0(\partial_t\bu+2\boldsymbol\Omega\times\bu)=&-\bn p+\rho_0\bn(\Phi+\Psi)+\rho\bn\Phi_0+\bF_\mathrm{\nu}+\bF_\mathrm{L},
\end{aligned}
\label{eq0}
\end{equation}
where $\rho_0$ is the mean density and  $\bu$, $p$, and $\rho$ are the perturbed velocity, pressure, and density,  respectively. We include the  volumetric  viscous force $\bF_\nu=\rho_0\nu\bd\bu$, which represents the effective  action of turbulent convection on tidal flows  with $\nu$ the effective so-called eddy-viscosity \citep[e.g.][]{Z1966b, Z1989, OL2012, MA2016} which we assume to be constant in the CZ. Moreover,
 \[
 \bF_\mathrm{L}=\frac{\bn\times\bB_0}{\mu_0}\times\bb+\frac{\bn\times\bb}{\mu_0}\times\bB_0
 \]
is the linearised Lorentz force, where $\bB_0$ and $\bb$ are the large-scale and perturbed magnetic fields,  respectively. Note  also that we have no background flow ($\bu_0=\bf 0$) as we work in the rotating frame, the action of the convective flows is parametrized as a diffusion, and any differential rotation and associated meridional flows are neglected.  Furthermore, the linearised  induction equation is:
\begin{equation}
    \partial_t\bb=\bn\times(\bu\times\bB_0)+\eta\bd\bb,
\label{ind}
\end{equation}
with $\eta$ the magnetic turbulent diffusivity, related to the eddy-viscosity by the relationship $\eta=\nu/\mathrm{Pm}$, where $\mathrm{Pm}$ is the turbulent magnetic Prandtl number often chosen close to unity \citep[e.g.][]{CT1992,JJ2013,KR2019}.

Following \cite{O2005,O2013}, we decompose all physical perturbed quantities $X$ into a non-wave like part associated with the equilibrium tide denoted  as $X_\h$, and a wave-like part $X_\mathrm{d}$ related to the dynamical tides. 
The equilibrium tidal flow $\bu_\h$ is defined as the velocity resulting from the hydrostatic adjustment of the primary due to the perturbation induced by the companion in the rotating frame of the tidal bulge.
This frame rotates at the corresponding tidal frequency $n \Omega_\mathrm{o}/2$ \citep[see][]{RM2012}, where $n$ labels the temporal harmonic of the orbital motion of the perturber (when projecting the tidal potential on the spherical harmonics basis) and $\Omega_\mathrm{o}$ is the associated orbital frequency.  In the adiabatic case, this hydrostatic equilibrium leads to \citep{Z1966a}:
\begin{equation}
-\bn p_\h+\rho_0\bn(\Phi_\h+\Psi)+\rho_\mathrm{e}\bn\Phi_0=0.
\label{eq0bis}
\end{equation}
This equation comes from Eq. (\ref{eq0}) without the left hand side  and dissipative terms. As a first step, we also  neglect the deformation of the stellar structure induced by rotation and magnetic field. 
In the case of magnetic fields this is a reasonable assumption except in the low density region near  the stellar surface \citep{DM2010}. 
 For the centrifugal acceleration, this is a fair hypothesis for slow and median rotators while potentially strong deformation should be taken into account for young rapid rotators \citep[see e.g.][Fig. 7]{GB2013}. 
We split  the equation of induction in the co-rotating frame, accounting for the equilibrium and dynamical tides decomposition:
\begin{subnumcases}{}
    \bb_\h=\bn\times(\boldsymbol\xi_\h\times\bB_0) \label{eq01a} \\
\partial_t\bb_\mathrm{d}=\bn\times(\bu_\mathrm{d}    \times\bB_0)+\eta\bd\bb_\mathrm{d}\label{eq01b}\ ,
\end{subnumcases}
where $\boldsymbol\xi_\h$ is  the equilibrium tide displacement, defined by $\bu_\h=\partial_t\boldsymbol\xi_\h$, given a  mean static magnetic field $\bB_0$. We also introduce $\bf u_\mathrm{d}$, the perturbed flow of the dynamical tide.  We assume here that $\bB_0$ does not vary over the tidal timescale which is a few days. 
This is corroborated by the fact that the large-scale magnetic field varies very little (far below one order of magnitude) within several years for the majority of observed stars \citep{VG2014}.  Moreover, we choose to define the magnetic field associated with  the equilibrium tide as the field it induces by the advection of $\bB_0$. We neglect the Ohmic diffusion acting on the equilibrium tide because its time scale $R^2/\eta$,  where $R$ is the radius of the star which is also the length of variation of this flow, is much larger than its typical time  of variation (in the range of a few days for Hot Jupiter), even when considering a turbulent magnetic diffusivity. One can not do the same assumption for dynamical tides since they involve potentially smaller length scales, for example along waves' attractors. The equation of induction for dynamical tides (Eq. (\ref{eq01b})) follows from this definition when writing  the equation of induction for the sum of the  equilibrium and dynamical tide  perturbations (Eq. (\ref{ind})) since it is a linear equation. 

In the momentum equation, we use the Cowling approximation \citep{C1941} for the  dynamical tides i.e. we neglect their  perturbed gravitational potential $\Phi_\mathrm{d}$. In addition, as our model applies to a convective region (i.e. adiabatically stratified), the term $\rho_\mathrm{d}\bn\Phi_0$ is neglected because it is related to the buoyancy force associated with gravity waves.  
 Using Eq. (\ref{eq0bis}), the momentum equation for tidally-forced magneto-inertial waves becomes:
\begin{equation}
\rho_0(\partial_t\bu_\mathrm{d}+2\boldsymbol\Omega\times\bu_\mathrm{d})+\bn p_\mathrm{d}-\bF_\nu(\bu_\mathrm{d})-\bF_\mathrm{L}(\bb_\mathrm{d})=f(\bu_\h),
\label{de_eq}
\end{equation}
where the wave-like part encompassing the propagation of tidal waves (on the left hand side of the equation), is excited by an effective force driven by  the equilibrium tidal flow (on the right hand side). This force  can be written as: 
\begin{equation*}
\begin{aligned}
f(\bu_\h)=&\underbrace{-\rho_0(\partial_t\bu_\h+2\boldsymbol\Omega\times\bu_\h}_{\fh})+\\
&\underbrace{\frac{\bn\times\bB_0}{\mu_0}\times\left[\bn\times(\boldsymbol\xi_\h\times\bB_0)\right]+\frac{\bn\times\left[\bn\times(\boldsymbol\xi_\h\times\bB_0)\right]}{\mu_0}\times\bB_0}_{\fm}.
\end{aligned}
\end{equation*}
 It is worth noting that the action of turbulent friction on  the hydrostatic flow has been neglected for the same reasons we ignored the ohmic diffusion in Eq. (\ref{eq01a}). 
The term $\fh$ comprises the driving inertial force and the Coriolis acceleration \citep[see the Appendix B in][]{O2005} while  $\fm$ embodies the action of the Lorentz force on the hydrostatic displacement and has been derived by \citet[see Appendix B]{LO2018}. These  authors studied the propagation and dissipation of magneto-inertial waves excited by the effective forcing induced solely by the Coriolis acceleration of the equilibrium tide (in short $f(\bu_\h)=\fh$). However, they also suggest that a large-scale magnetic field can potentially interact with the equilibrium tide for sufficiently large Lehnert numbers (typically $\Le>0.1$). For this reason, we propose to examine the relative importance of both forcings  $\fm$ and $\fh$.
We use $R$ as \bfo the typical length scale of the large-scale magnetic field and of \bfc the equilibrium tide, which involves large-scale flows. Henceforth, we can give the order of magnitude of the different forcings:
\begin{equation*}
\left\{\begin{aligned}
      \fh&\sim \rho_0u_\h\sm\hspace{0.5
      cm}\text{ with }\sigma_\mathrm{max}=\max{[\st,2\Omega]}\\
      \fm&\sim B_0^2\xi_\h/(\mu_0 R^2)
\end{aligned}\right.,
\end{equation*}
and their ratio:
\begin{equation}
\frac{\fm}{\fh}\simeq\frac{B_0^2}{\rho_0\mu_0 (2\Omega R)^2}\times\frac{2\Omega\xi_\h}{u_\h}\times\frac{2\Omega}{\sm}\equiv\Le^2\times\Roh^{-1}\times\scm^{-1},
\label{eq1}
\end{equation}
where we define the Lehnert number as $\Le=B_0/(\sqrt{\rho_0\mu_0}2\Omega R)$. We also introduce the Doppler-shifted tidal  Rossby number $\Roh=\st/(2\Omega)$ with $\st=u_\h/\xi_\h$ the related tidal frequency, and a dimensionless frequency ratio $\scm=\sm/(2\Omega)=\max[\Roh,1]$. 
 According to Eq. (\ref{eq1}), magnetism needs to be taken into account  in the tidal forcing whenever $\Le^2/(\Roh\scm)\gtrsim1$.
\subsection{Scaling laws to estimate stellar magnetic fields}
\label{sec22}
The determination of the Lehnert number inside the convective envelope of low-mass stars requires knowledge of the internal magnetic field of these stars. 
However, we are currently  only able to constrain the magnetic field of stars  at their surface. Indeed, thanks to Zeeman-Doppler Imaging, one can reconstruct the topology and strength of large-scale, stellar magnetic fields  \citep{D2006,DJ2007}.

Regarding the Sun, internal magnetic fields can be assessed indirectly from surface tracers like sunspots, as a manifestation of flux ropes emerging through  the surface \citep{C2013}. 
This approach is based on the so-called interface dynamo theory where the magnetic field is generated by a convective dynamo and pumped into the tachocline  (the interface between the radiative and convective zones), where it eventually becomes strong enough to be buoyantly unstable, rise through the convection zone, and emerge at the surface as sunspots  \citep{SZ1992,C2014,BB2017}. This picture  has been recently questioned, especially by \cite{WD2016} by  studying fully convective stars. It is indeed possible that the spot-forming magnetic fields can be generated in the bulk \citep{SB2017} or in the shallow layers \citep{B2016} of the CZ instead of overshoot layers beneath the core-envelope interface.
Through helioseismology, \cite{GT1990} and \cite{A2000} have placed an upper bound of 30 T on the toroidal magnetic field strength at the base of the CZ.
On the contrary, the mean magnetic field at the Sun's  surface is significantly smaller, about a few Gauss  ($\sim 10^{-4}\,\tesla$), even though sunspots are the seats of local intense magnetic fields (several tenths of a Tesla).

As far as younger or smaller stars than the Sun are concerned, \cite{VG2014} gives  an overview  of known large-scale surface magnetic fields, that appear to vary  from the globally weak fields  of the order of  the Gauss for solar-like stars to the strong Tesla-strength fields of  M dwarf and T-Tauri stars. They establish  a relationship between surface magnetic field and the convective Rossby number $\Ro=\uc/(2\Omega\lc)$, where $\uc$ and $\lc$ are the convective velocity and length, respectively. This dimensionless quantity is generally  calculated at half the mixing-length $\alpha H_p/2$ \citep{LM2010,G1986} above the base of the CZ, with $H_p$ the pressure scale height and $\alpha$ the mixing-length theory coefficient.

Nonetheless, as indicated above, it is difficult to estimate the magnetic field inside stars.  
In this context, 3D global non-linear  simulations \citep{SB2017,EB2017} and scaling relationships for stellar dynamos \citep[ and references therein]{AB2019} can help us to estimate  the internal magnetic field strength in the convective envelope of  low-mass stars.
Three scaling laws are described hereafter and the derivation of the related dynamo-induced magnetic field is detailed in Appendix \ref{AA}. We have:
\begin{enumerate}
\item \textit{the  (turbulent) equipartition}, that is often  used to give  an averaged,  rough estimate of the magnetic field's amplitude in the bulk of the CZ \citep{BB2017}. It assumes that the dynamo is efficient i.e. that the system is equally good at generating magnetic field as it is at generating flows.
This balance is also used in moderately active plages at the solar surface, while in the active sunspots superequipartition (i.e. magnetic energy is greater than kinetic energy) can be fairly common \citep{DL2009}.
\label{item1}
\begin{table}[t]
\centering
\begin{threeparttable}
\begin{tabular}{lll}
Regime & Balance & Estimation of $B_\mathrm{dyn}$ \\
\hline\hline
\rule[-2ex]{0pt}{5ex} \ref{item1} Equipartition & $\ME=\KE$ & $\sqrt{\mu_0 2\KE}$\\
\rule[-2ex]{0pt}{5ex} \ref{item2} Buoyancy dynamo 
& $\cfrac{\mathrm{ME}}{\mathrm{KE}}=\Ro^{-1/2}$ & $\sqrt{\mu_02\KE/\Ro^{1/2}}$\\
\rule[-2ex]{0pt}{5ex} \ref{item3} Magnetostrophy & ${\bf F}_\mathrm{L}=2\rho_0\boldsymbol\Omega\times\bu$ & $\sqrt{\mu_02\KE/\Ro}$ \\
\end{tabular}
\end{threeparttable}
\caption{Magnetic fields derived from simple balances (forces or energies) written in the second column.
Details of the calculations are given  in the appendix \ref{AA}. $\KE$ and $\ME$ are the kinetic and magnetic energy densities of the convective flow, respectively.}
\label{tab1}
\end{table}
\item \textit{the  buoyancy dynamo  regime}, in which the  Coriolis, buoyancy and Lorentz forces are taken to have the same order of magnitude  assuming a low atomic  magnetic Prandtl number, that is the ratio of atomic viscosity to magnetic diffusivity  \citep{D2013,AB2019} . This assumption is well verified for fast rotating giant planets, young contracting stars like T-Tauri and rapidly-rotating low-mass stars \citep{CH2009}. 
\label{item2}
\item \textit{the  magnetostrophic regime}, for which  the force balance is realised  between the Coriolis and Lorentz forces. This balance,  also called magnetostrophy, gives an upper estimate of the magnetic field. The magnetostrophic  regime generally gives fields in super-equipartition \citep{BG2015,AB2019}.
\label{item3}
\end{enumerate}
We summarise in Table \ref{tab1} the order of magnitude estimation of the magnetic field  (named hereafter $B_\mathrm{dyn}$) within the convective envelope of a low-mass star as obtained by the three aforementioned scaling laws.  These relationships involve the kinetic and magnetic energy densities of the convective flow $\KE=\rho\uc^2/2$ and $\ME=B_\mathrm{dyn}^2/(2\mu_0)$ respectively, along with the convective Rossby number $\Ro=\uc/(2\Omega\lc)$ at the base of the CZ. \\

\bfo The main objective of this paper is to determine and quantify the impact of a large-scale magnetic field on the excitation and dissipation of the dynamical tides. We still discuss in Sect. \ref{sec4} to what extent the small-scale component of the stellar magnetic field can influence the tidal flows. \bfc
To allow comparison with the amplitude of observed surface magnetic fields of low-mass stars, we extrapolate a \bfo large-scale, dipolar, surface \bfc magnetic field from the scaling laws at the base of the CZ (Table \ref{tab1}). First, we suppose that the dipolar component of the star's  magnetic field at the base of the CZ is a fraction $\gamma$ of the dynamo-induced magnetic field at that location:
\begin{equation}
B_\mathrm{dip}(r_\mathrm{base})=\gamma B_\mathrm{dyn}(r_\mathrm{base}).
\label{gamma}
\end{equation}
The factor $\gamma$ encapsulates both the ratio of large-scale to small-scale magnetic fields \cite[similarly to the filling factor in][]{R2012,SM2019} and the part of the total energy that is available in the dipolar component of the magnetic field.

Then, we infer the dipolar component of the surface magnetic field simply as:
\begin{equation}
B_\mathrm{dip}(r_\mathrm{top})=(r_\mathrm{base}/R)^3B_\mathrm{dip}(r_\mathrm{base}),
\label{extrapol}
\end{equation}
where $R$ is the radius of the star, the subscripts "base" and "top" are the position inside the CZ, and  "dip" and "dyn" refer to the dipolar and dynamo-induced magnetic fields, respectively. As the dipolar magnetic field at the surface of the Sun  is well known, typically $4\,\mathrm{G}$ \citep{DB2012}, $\gamma$ can be estimated for the Sun by using Eqs. (\ref{gamma}) and (\ref{extrapol}): 
\begin{equation}
\gamma=\frac{B_{\mathrm{dip},\odot}(r_\mathrm{top})}{ B_{\mathrm{dyn},\odot}(r_\mathrm{base})}\times\left(\frac{r_\mathrm{base,\odot}}{R_\odot}\right)^{-3},
\label{gam2}
\end{equation}
with $R_\odot$, $B_{\mathrm{dyn},\odot}(r_\mathrm{\bf base})$ and $r_\mathrm{\bf base,\odot}$ obtained from the grid models of the STAREVOL evolution code (see the Sect. \ref{sec23} below) for a $1M_\odot$ star of the age of the Sun. In the following we will assume that $\gamma$ in Eq. (\ref{gamma})  is independent of the mass and the age of the star, and that it takes the Sun's current value as in Eq. (\ref{gam2}). This is a strong assumption since $\gamma$ is close to a filling factor that depends on the Rossby number and therefore on the angular frequency of the star \citep{SM2019}. Nevertheless, refining the expression of this factor would not change the final conclusions of this paper which are robust to several orders of magnitude, as we will see later.
The factor $\gamma$ however depends on the scaling law used to estimate $B_{\mathrm{dyn},\odot}$.

Finally, the dipolar component of a star's surface magnetic field will be estimated as 
\begin{equation}
B_\mathrm{dip}(r_\mathrm{top})=\gamma (r_\mathrm{base}/R)^{3} B_\mathrm{dyn}(r_\mathrm{base}).
\label{Bdip}
\end{equation}
\subsection{The stellar evolution code  STAREVOL}
\label{sec23}
To estimate the magnetic field and then the Lehnert number in the convective envelope of low-mass stars of different ages via the scaling laws described in Sect \ref{sec22}, we use  the 1D stellar evolution code STAREVOL \citep{AP2019}. Initial masses of the stars range from $0.4$ to $1.4M_\odot$, given a solar metallicity $Z=0.0134$ \citep{AG2009}, and a mixing length parameter $\alpha=1.9730$. This latter is  defined by the calibration of the standard solar model and used to model convective regions according to the mixing length theory. 
Basic input microphysics like the equation of state, nuclear reactions or opacities, are described in \cite{AP2016} and  \cite{LD2012}. 
The initial rotation periods are fixed using the calibration for fast ($1.6$ days), median ($4.5$ days) and slow ($9$ days) rotators from \cite{AP2019}.
The rotation is assumed to be uniform  inside the CZ but varies dramatically with  stellar ages \citep{GB2013}. As a result,  the evolution of the surface  angular velocity dictates that of  the Lehnert number. 
During the first few Myr of the pre-main sequence (PMS), the surface angular velocity of the stars remains stationary  as the result of star-disk magnetic interactions  \citep{ZF2013,GB2015,AP2016}. This holds over  the disk lifetime, typically a few Myr  \citep{RW2004,BN2013,GB2015}. 
 After the dissipation of the  disk, the gravitational contraction of the star  leads to an increase in the angular velocity, in order to conserve angular momentum, until the star begins hydrogen fusion at the zero-age main sequence (ZAMS). From this stage onward,  magnetised stellar winds apply a torque on the star, spinning it down throughout its  main sequence (MS) lifetime. In the STAREVOL code, the effects of a stellar wind acting from the early PMS to the tip of the MS are  implemented using the prescription given by  \cite{MB2015}.
\begin{figure*}[ht]
\includegraphics[width=\hsize/2]{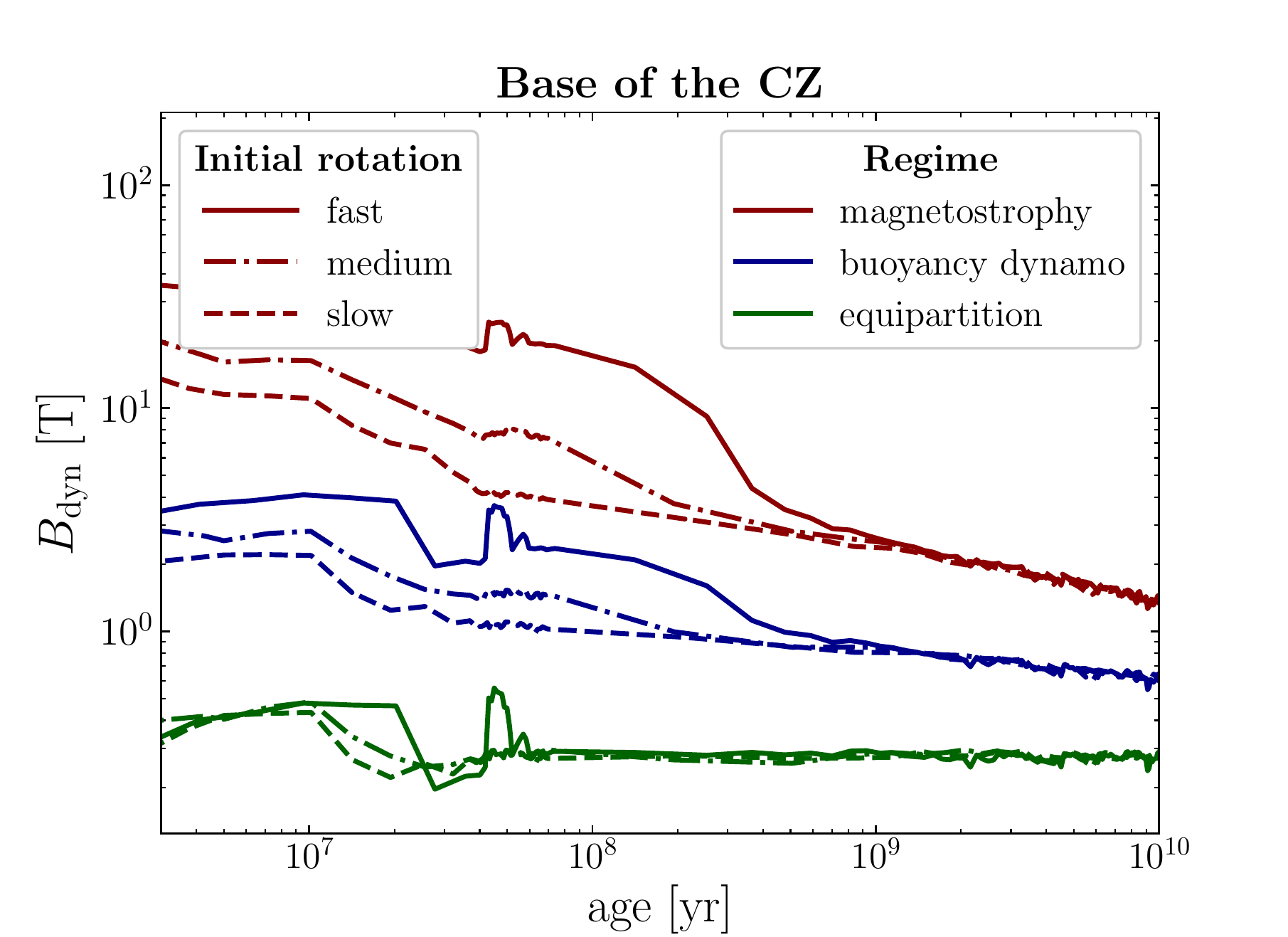}
\includegraphics[width=\hsize/2]{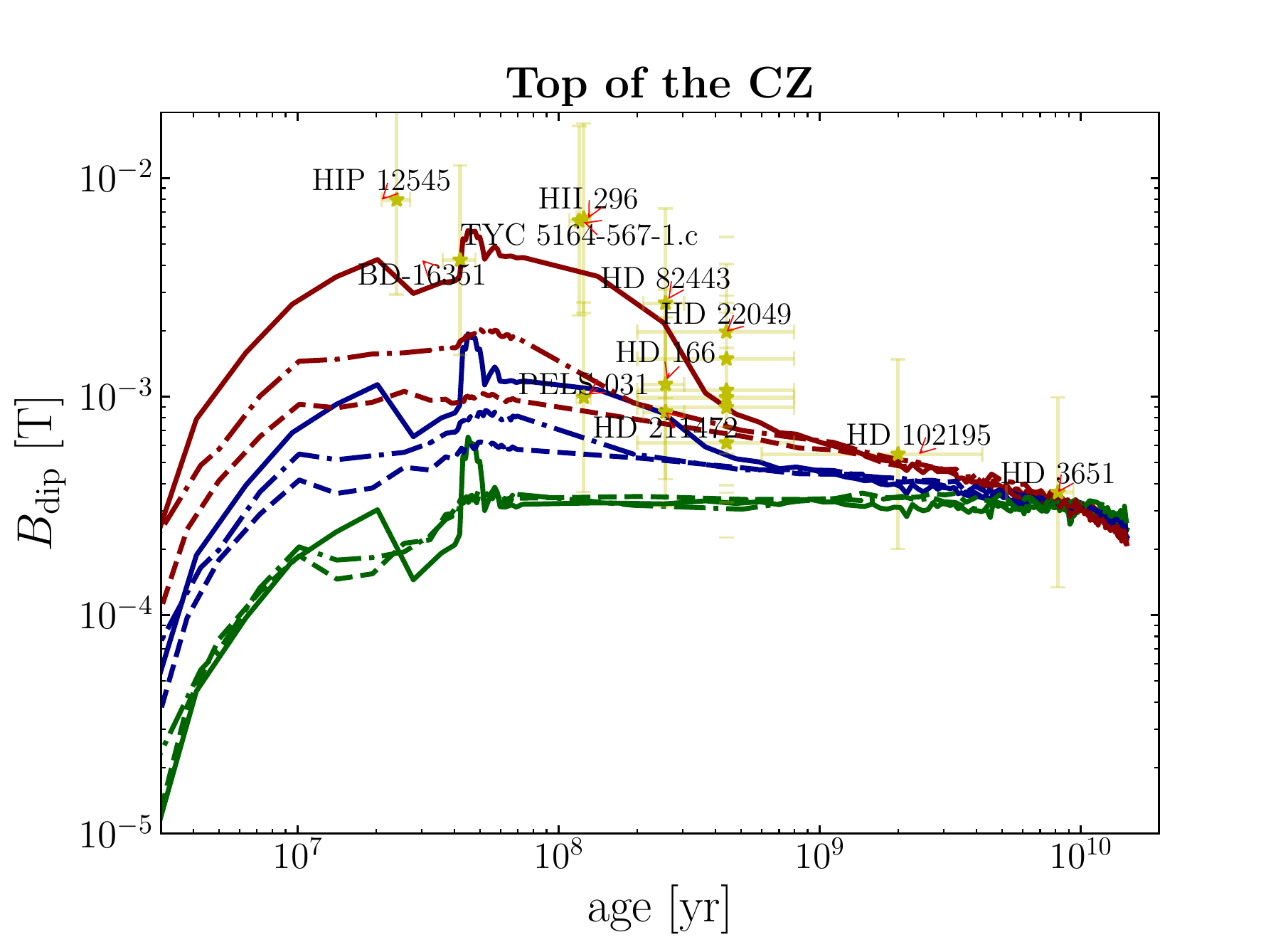}
\caption{Magnetic field versus the age of a $0.9M_\odot$ star for three different initial rotation periods. At the base of the convective zone (\textit{left panel}), the magnetic field is estimated from the scaling laws listed in Table \ref{tab1}. At the top of the convective zone (\textit{right panel}), the same scaling laws are used to extrapolate a dipolar magnetic field near the surface (see Eq. (\ref{Bdip})).
The "stars" symbols {\color{jv}\ding{72}} that represent mean dipolar magnetic fields at the surface of $0.9M_\odot$ stars are taken from \cite{SJ2017}, and the age of the stars are from \cite{GE2006,JR2008,LT2015} and \cite{FP2016}. Multiple observations of the dipolar magnetic field of a star are joined by a line. }
\label{fig1}
\end{figure*}
In the following, the base of the CZ refers to a height that is located  $0.002R$ above the bottom of the CZ as computed  in the STAREVOL model. This convention  avoids erratic numerical behaviour of the mixing length  convective velocity  at the interface between the radiative and convective zones. Furthermore, the top of the CZ refers to the radius where the convective velocity vanishes in the STAREVOL models.
\subsection{Estimation of the dipolar magnetic field at the base and the top of a convective zone}
\label{sec24}
Figure \ref{fig1} shows the evolution of the magnetic field of a $0.9M_\odot$ star along its lifetime for the three different initial rotation rates  stated in Sect. \ref{sec23}. It should be specified that we chose a $0.9M_\odot$ star rather than $1M_\odot$ to add measurements of the mean dipolar magnetic fields at the surface of stars in the early MS (see next paragraph).  
At the base of the CZ (left panel), the magnetic field is calculated with the scaling laws listed  in Table \ref{tab1}. At the top of the CZ (right panel), we use these dynamo-induced magnetic fields to extrapolate dipolar magnetic fields near the surface by using Eq. (\ref{Bdip}). The results obtained with  fast ($1.6\,\mathrm{ days}$), median  ($4.5\,\mathrm{ days}$), and slow ($9\,\mathrm{ days}$) initial rotations are plotted with  solid, dash-dotted, and dashed curves,  respectively. We note that the magnetic field decreases in time  after about $50$ Myr in the two panels for both the  magnetostrophic and buoyancy dynamo  regimes. It is due to the fact that these regimes depend on a positive power of  the angular velocity  (in the denominator of the Rossby number, see Table \ref{tab1}) that decreases after the ZAMS as a result of the stellar wind action on the star's surface. After $1\,\mathrm{Gyr}$, this decline is well described by the empirical Skumanich relationship \citep{WD1967,S1972}.
At the age of the present Sun ($\sim 4.6\,\giga\mathrm{yr}$), the buoyancy dynamo and magnetostrophic regime at the base of the CZ give the order of magnitude of the toroidal magnetic field strength expected in the Sun at the tachocline, typically a few to a few tens  of Tesla \citep{C2013}.

We have added on the plot for the top of the CZ (right-hand panel)  the  average unsigned measured dipolar field strength of $0.9M_\odot$ stars \citep{SJ2017}. We have adopted a  conservative error estimate of 0.434 dex in $\log B_\mathrm{dip}$ for all stars. Note that HD 22049 displays several values of the amplitude of the dipolar magnetic field measured at different times.  The ages of the stars and their errors are taken from \cite{FP2016} for the early MS stars and \cite{GE2006, JR2008,LT2015} for the three oldest stars.
It is interesting to note that the strength of the observed dipolar magnetic fields seems to be steady for end PMS and early MS stars. This finding is consistent with the saturation levels observed by \cite{VG2014,SJ2017} for the large-scale and dipolar surface magnetic fields at low Rossby numbers. As a matter of fact, low Rossby numbers means high angular frequencies and therefore stars in the vicinity of the ZAMS.
However, whether or not this observed saturation level is an atmospheric effect or a dynamo related phenomenon has yet to be determined \citep{V1984,JU1999}. 

Dipolar fields  of young stars seem to fit quite well with the magnetostrophic regime along  with fast initial rotation. We must nevertheless specify that we realised the same plot with more observed dipolar magnetic fields (See et al. in preparation, private communication) featuring weaker magnetic fields for PMS and early MS stars, and so consistent with median and slow initial rotation.
That being said, we have to bear in mind that several strong assumptions have been made to derive the dipolar magnetic field near the surface (Eq. (\ref{Bdip}), see also Appendix \ref{dis}). These prescriptions are nonetheless sufficient for this work given the robustness of our results as we will see in Sect. \ref{sec26}. 

\begin{figure}[th]
\includegraphics[trim=0cm 0cm 1cm 0cm clip,width=\textwidth/2]{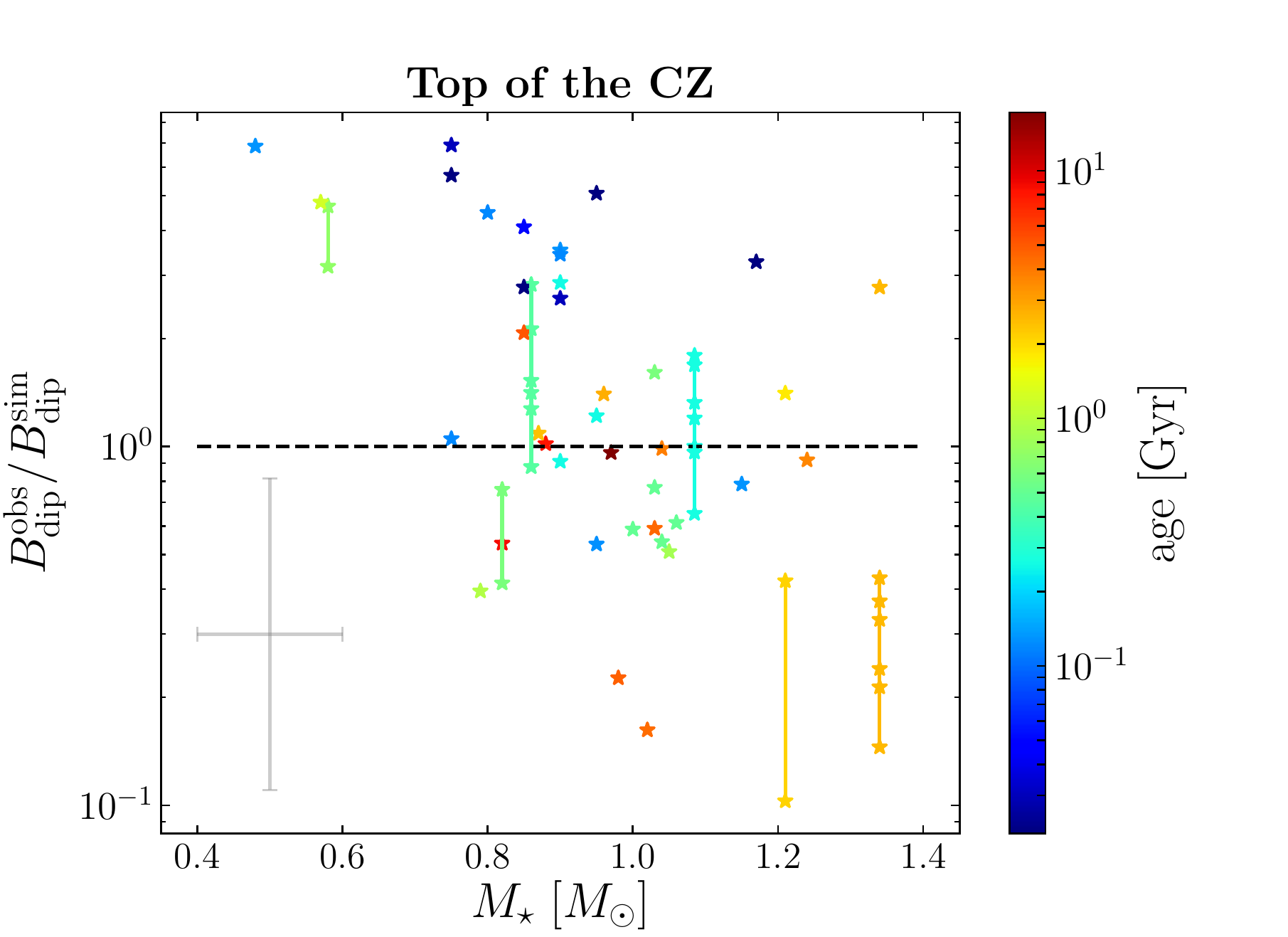}
\caption{Ratio of the mean unsigned observed  dipolar magnetic field $B_\mathrm{dip}^\mathrm{obs}$ \citep{SJ2017,VG2014} over estimated dipolar magnetic field $B_\mathrm{dip}^\mathrm{sim}$ as a function of  the mass of the star. The age of the stars is indicated in colour and ranges from $21$ Myr to $17.2$ Gyr.  Magnetostrophic regime and median initial rotation have been  used to plot the ratio $B_\mathrm{dip}^\mathrm{obs}/B_\mathrm{dip}^\mathrm{sim}$. Multiple observations of the dipolar magnetic field of a star are again  joined by a line. A typical error bar for $M_\star$ and $B_\mathrm{dip}^\mathrm{obs}/B_\mathrm{dip}^\mathrm{sim}$ is indicated in grey. We have adopted a conservative error estimate of 0.434 dex in $\log B_\mathrm{dip}$ and $0.1M_\odot$ in $M_\star$.}
\label{BobsBdyn}
\end{figure}

We have plotted in Fig. \ref{BobsBdyn} the ratio of the observed and estimated dipolar magnetic fields versus the mass of various low-mass stars distributed from PMS to MS stages. The magnetic field $B_\mathrm{sim}$ is calculated with the magnetostrophic regime and a median initial rotation. The magnetic field $B_\mathrm{obs}$ is again taken from \cite{SJ2017} and ages are from \cite{VG2014} based on different methods (see the last quoted paper for more details). We note that the surface dipolar magnetic field of stars is well reproduced by $B_\mathrm{sim}$ within an order of magnitude. 
The estimate $B^\mathrm{sim}_\mathrm{dip}$ tends to deviate from $B^\mathrm{obs}_\mathrm{dip}$ for relatively massive ($M_\star\gtrsim1.2M_\odot$) or very low-mass ($M_\star\lesssim 0.6M_\odot$) stars, which is not surprising considering that the ratio $\gamma$ of dipolar/non dipolar magnetic field is determined from solar parameters (see Eq. (\ref{gam2})). 

\subsection{Lehnert number for a low-mass star along its evolution}
\label{sec25}
Using the estimates  of the dynamo-induced magnetic field in Table \ref{tab1}, we can express  the related Lehnert number at the base of the CZ  (see Table \ref{tab2}). Within each regime, $\Le$ depends on the convective Rossby number, and the ratio between the  convective length scale  and the  stellar radius $\lc/R$. This ratio results  from the different length scales used in the magnetic scaling laws and  in the definition of the Lehnert number (the length scales $\lc$ and $R$ respectively). Specifically,  we choose a definition of the Lehnert number that  is consistent with the previous works of \citet{LO2018} and \cite{W2018}. 
\begin{table}[h]
\centering
\begin{threeparttable}
\begin{tabular}{cc}
Regime & $\Le$ scaling  \\
\hline\hline
\rule[-2ex]{0pt}{5ex} Equipartition & $\Ro\times\lc/R$\\
\rule[-2ex]{0pt}{5ex}Magnetostrophy & $\sqrt{\Ro}\times\lc/R$ \\
\rule[-2ex]{0pt}{5ex} Buoyancy dynamo & $\Ro^{3/4}\times\lc/R$\\
\end{tabular}
\end{threeparttable}
\caption{Lehnert number scaling laws depending on magnetic field regimes, detailed in Table \ref{tab1}.}
\label{tab2}
\end{table}

At the top of the CZ, we use the dipolar magnetic field derived in Eq. (\ref{Bdip}) to estimate the Lehnert number near the surface:
\begin{equation}
\Le_\mathrm{top}=\frac{B_\mathrm{dip}(r_\mathrm{top})}{\sqrt{\mu_0\rho_\mathrm{top}}2\Omega R}=\gamma\left(\frac{r_\mathrm{base}}{R}\right)^3\sqrt{\frac{\rho_\mathrm{base}}{\rho_\mathrm{top}}}\Le_\mathrm{base},
\label{Letop}
\end{equation}
where we recall that "top" and "base" refer to the position in the CZ and that $\Le_\mathrm{base}$ is taken from Table \ref{tab2}.
\begin{figure*}[h]
\resizebox{\hsize/2}{!}{\includegraphics{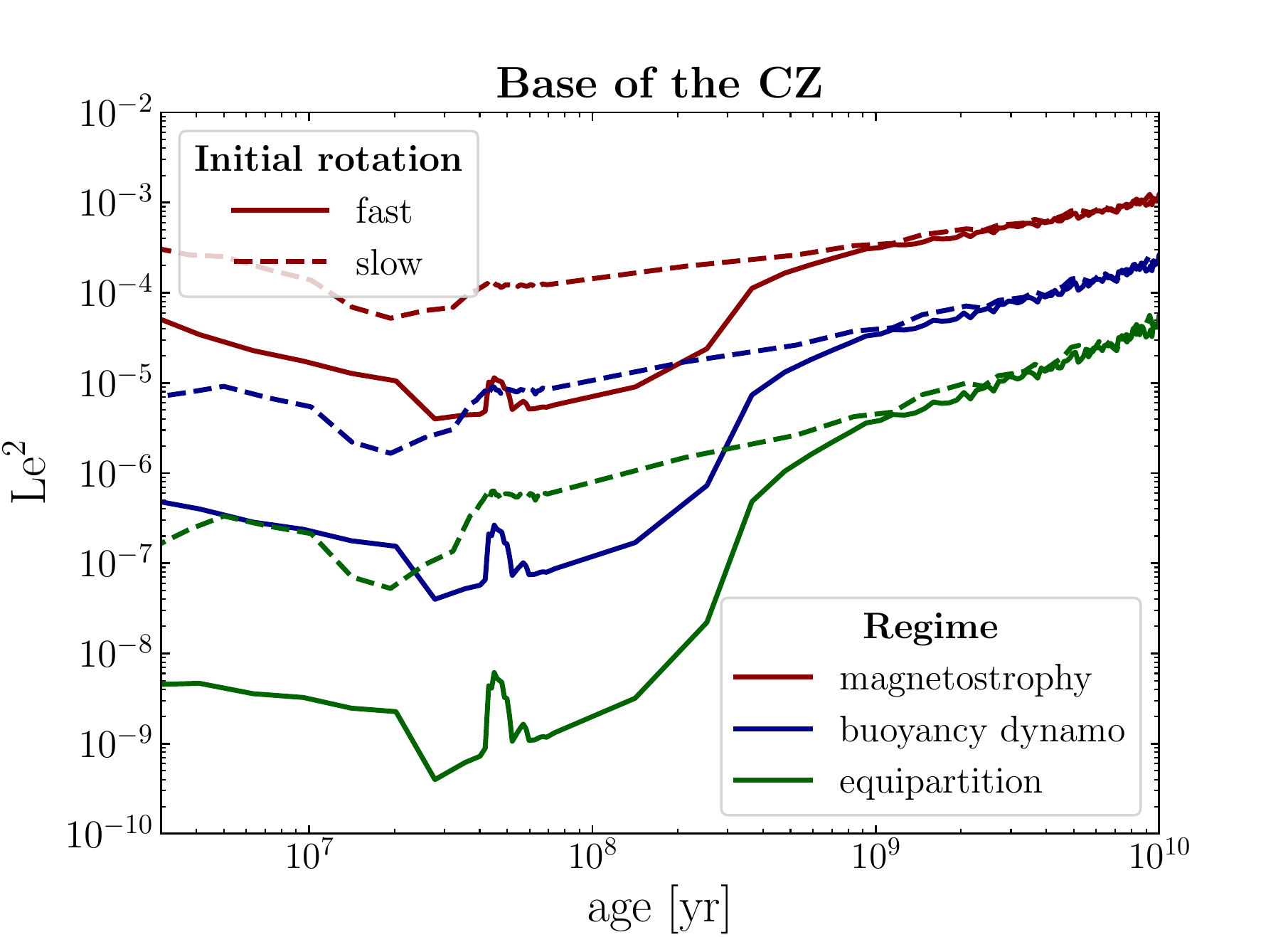}}
\resizebox{\hsize/2}{!}{\includegraphics{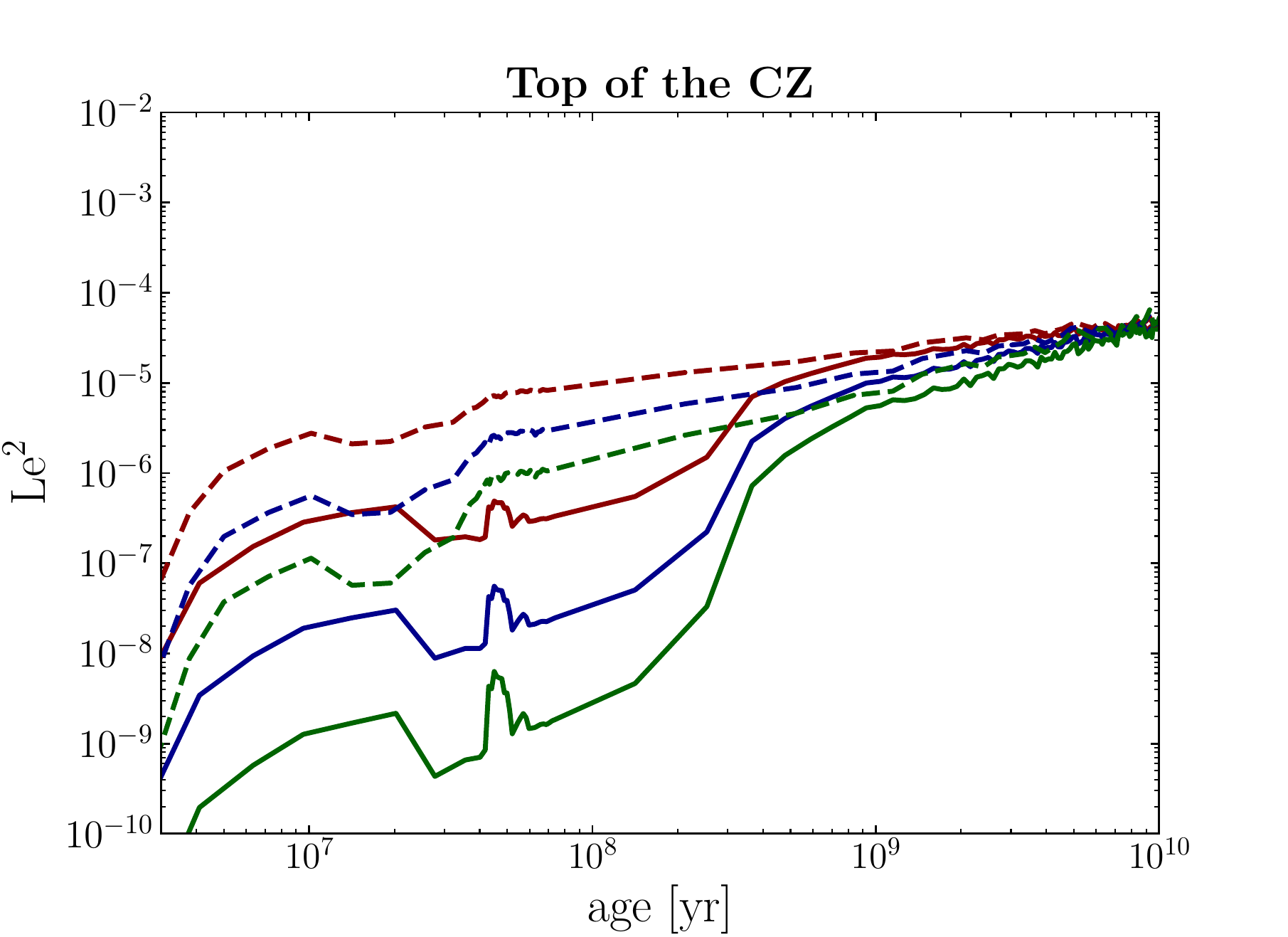}}
\caption{Lehnert number squared at the base (\textit{left panel}) and the top (\textit{right panel}) of the convective zone as a function of age for a $1M_\odot$ star, and for different magnetic scaling laws and initial rotations  (see the legends in the left panel).}
\label{fig2}
\end{figure*}

In Fig. \ref{fig2}, the Lehnert number squared of a $1M_\odot$ star is shown as a function of age, at the base (left panel, see Table \ref{tab2}) and top (right panel, Eq. (\ref{Letop})) of the CZ. 
The different scaling laws, with fast and slow initial rotations, are shown with the same layout as Fig. \ref{fig1}.  
First, we note that $\Le^2$ remains always smaller than unity, consistent with the previous  works of \citet{LO2018} and \cite{W2018}. Whether we look at the base or the top, $\Le^2$ increases with decreasing initial rotation speed  from the PMS until about $1\mathrm{Gyr}$. This is the opposite behaviour to  the magnetic field strength,   since the Lehnert number decreases with  the angular velocity (see Table \ref{tab2} and Eq. (\ref{Letop})). Finally,  $\Le^2$ is greater  at the base than at the  top of the CZ as it scales with $B^2$.
\begin{figure}[t]
\resizebox{\hsize}{!}{\includegraphics{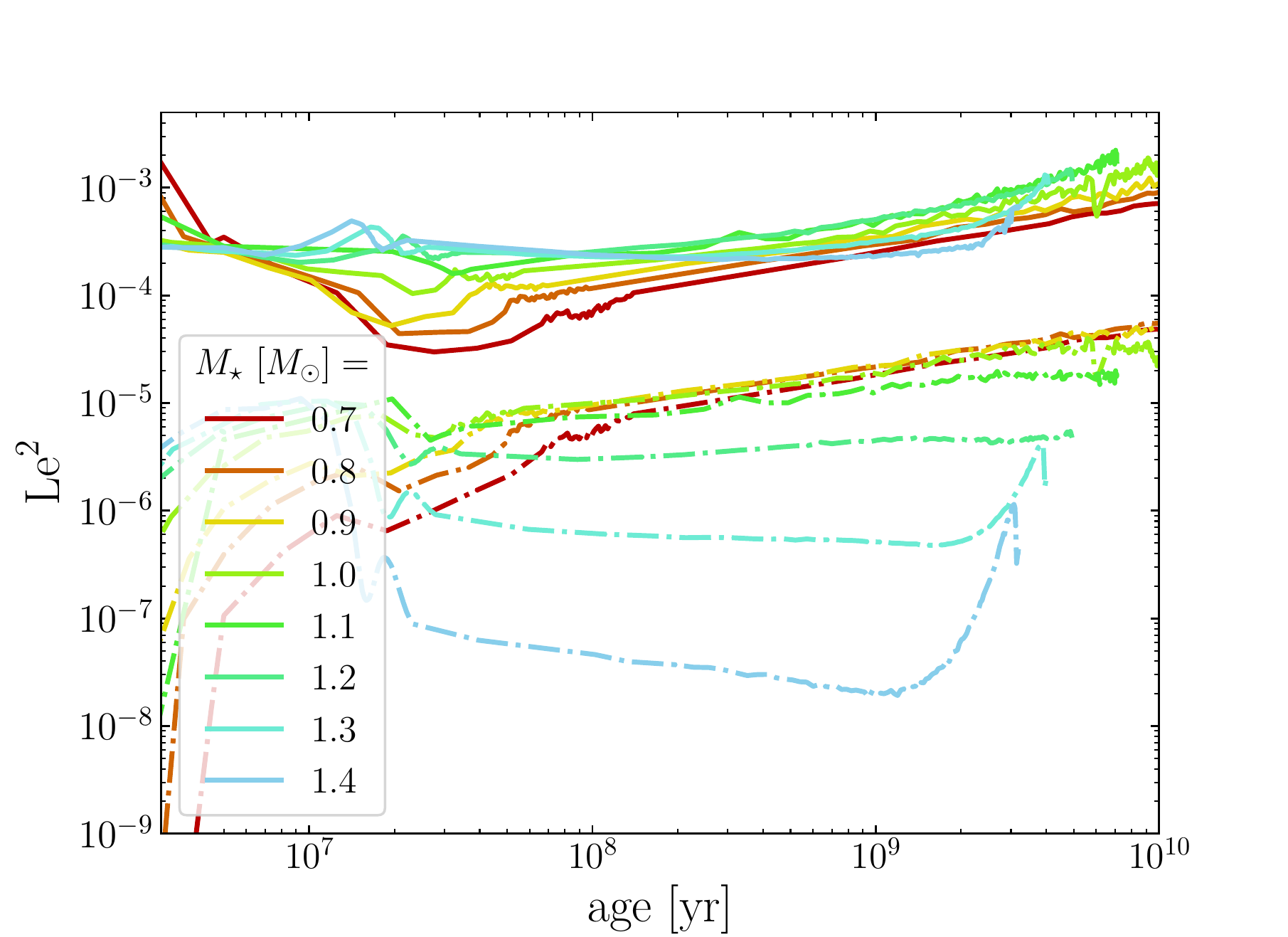}}
\caption{Evolution of the  Lehnert number squared over time, at the base (solid lines) and the top (dotted dashed lines) of the convective zone,  for various low-mass stars sorted from $0.7$ to $1.4M_\odot$. Slow initial rotation has been used here and a  magnetostrophic balance has been assumed.}
\label{fig3}
\end{figure}

The mass dependence of the Lehnert number is illustrated   in Fig. \ref{fig3}, which  displays $\Le^2$ against age for stars ranging from $0.7$ to $1.4M_\odot$. We have used slow initial rotation and the magnetostrophic regime to plot $\Le^2$ at both  the base and the top of the CZ. 
At the base of the CZ, we obtain the same features as in Fig. \ref{fig2} i.e. $\Le^2$ reaches a minimum at the ZAMS and  increases during the MS. We highlight a quite  small dispersion of the curves, though $\Le^2$ still grows slightly with mass for $M_\star\lesssim1.2\,M_\odot$ stars. This trend is  driven mostly by the decay of the convective turnover time $\tau_\mathrm{c}=\lc/\uc$ when the mass grows (note that $\Ro=(\tau_\mathrm{c}2\Omega)^{-1}$ in Table \ref{tab2}). When $M_\star\gtrsim1.2\,M_\odot$, the decrease in $\lc/R$ at the base of the CZ for growing masses helps to reverse this behaviour.
The drop in $\Le^2$ with mass is even more pronounced at the top of the CZ for $M_\star\gtrsim1\,M_\odot$.
These changes simply reflects the fact that the convective envelope shrinks with mass, which naturally leads to an increase in convective length and a decrease in convective turnover time and density at the base of the CZ (see Table \ref{tab2} and Eq. (\ref{Letop})).
\subsection{The influence of magnetism on tidal forcing for observed star-planet systems}
\label{sec26}
The ratio of the magnitude of the  Lorentz forcing to the  hydrodynamical forcing (Eq. (\ref{eq1})) does not only depend on the Lehnert number, but also on the Doppler-shifted Rossby number $\Roh$ and the ratio of frequencies $\scm$ which need to be estimated from a two-body system.
\cite{O2014} gives the expression of the tidal frequency in the fluid frame, $\st$ in our  notations, using  the integers $l$, $m$, $n$ coming from the spherical harmonics functions on which the gravitational potential is projected \citep[we refer the reader to Sect. 2.1 in][]{O2014}. The tidal frequency is $\st=n\Oo-m\Omega$, where $m$ is the azimuthal  order of the spherical harmonics, and $n$ and $\Omega_\mathrm{o}$ have been  introduced in Sect.  \ref{sec21}. 
In the previous section, it has been shown that $\Le^2$ does not go much beyond $10^{-3}$, regardless of the age and the mass of the star at the base and top of the CZ (Figs. \ref{fig2} and \ref{fig3}).  This means that the ratio $\fm/\fh$ (Eq. (\ref{eq1})) is likely to be small compared to unity, unless the rotation frequency is much greater (by at least a factor of a thousand) than the tidal frequency. Given the definition of the tidal frequency introduced above, the closer we get to a resonance between orbital and rotation frequencies, the more important the ratio $\fm/\fh$ will be.

For the sake of simplicity we will consider here systems with an almost circular and coplanar orbit.
This allows us to reduce the number of pairs $(m,n)$ because the tidal potential components are labelled by these integers and depend on eccentricity and stellar  obliquity.
Moreover, the dominant term in the tidal potential is the quadrupolar component as long as the planet and its host star are well separated, namely $l=2$ with $l$ the degree of the spherical harmonics \citep{ML2009,O2014}. 
Within this assumption and the limits of low eccentricity and obliquity,  $(m,n)\in\{(2,2),(0,1),(2,1),(2,3),(1,0),(1,2)\}$ \citep[see][for more mathematical  details]{O2014}. When the orbit is strictly circularised and coplanar, the asynchronous tide acts alone and the only matching pair of integers  is $(2,2)$. For the other pairs, the eccentricity or obliquity tides can be dominant. \\

In Table \ref{tabsys}, we have thus  listed known star-planet systems satisfying the following conditions:
\begin{itemize}
\item a near-circular orbit: we choose  the eccentricity such as  $e<0.1$.
\item a low sky-projected obliquity: $|\lambda|<30\degree$. $\lambda$ is the sky-projected angle between the stellar spin axis and the axis perpendicular to the planet orbit.   Ideally, we should use the true  obliquity $\psi$   but this quantity is more difficult to determine than $\lambda$ and too few measurements exist. However, when both values of  $\psi$ and $\lambda$ exist for the selected systems,  they are quite similar and far from the threshold of $30\degree$.
\item a planet orbiting close to its host star : $P_\mathrm{o}<10$ days, where $P_\mathrm{o}$ is the orbital period of the planet, so that stellar tidal effects are important .
\end{itemize}
Under these conditions the chosen systems are mostly hot-Jupiter-like systems, with host stars ranging from $0.7$ to $1.4\,M_\odot$ in order to have a similar structure to that of the Sun (namely with a convective envelope and a radiative region below it during the MS). They have been picked out using the online database
\href{http://www.exoplanet.eu}{exoplanet.eu}\footnote{http://www.exoplanet.eu} \citep[e.g.][]{SD2011}  
which presents the  orbital period and eccentricity of the planet, along with the \href{http://www.astro.keele.ac.uk/jkt/tepcat/tepcat.html}{TEPcat}\footnote{www.astro.keele.ac.uk/jkt/tepcat/tepcat.html} \citep{S2011} database  to find the sky-projected obliquity. 
Then, we removed the systems for which the age of the star was not known. 
 For the remaining systems, stellar period has been found in the  literature: the related references are reported in the last column of Table \ref{tabsys}.

\begin{figure}[ht]
	\includegraphics[width=\textwidth/2]{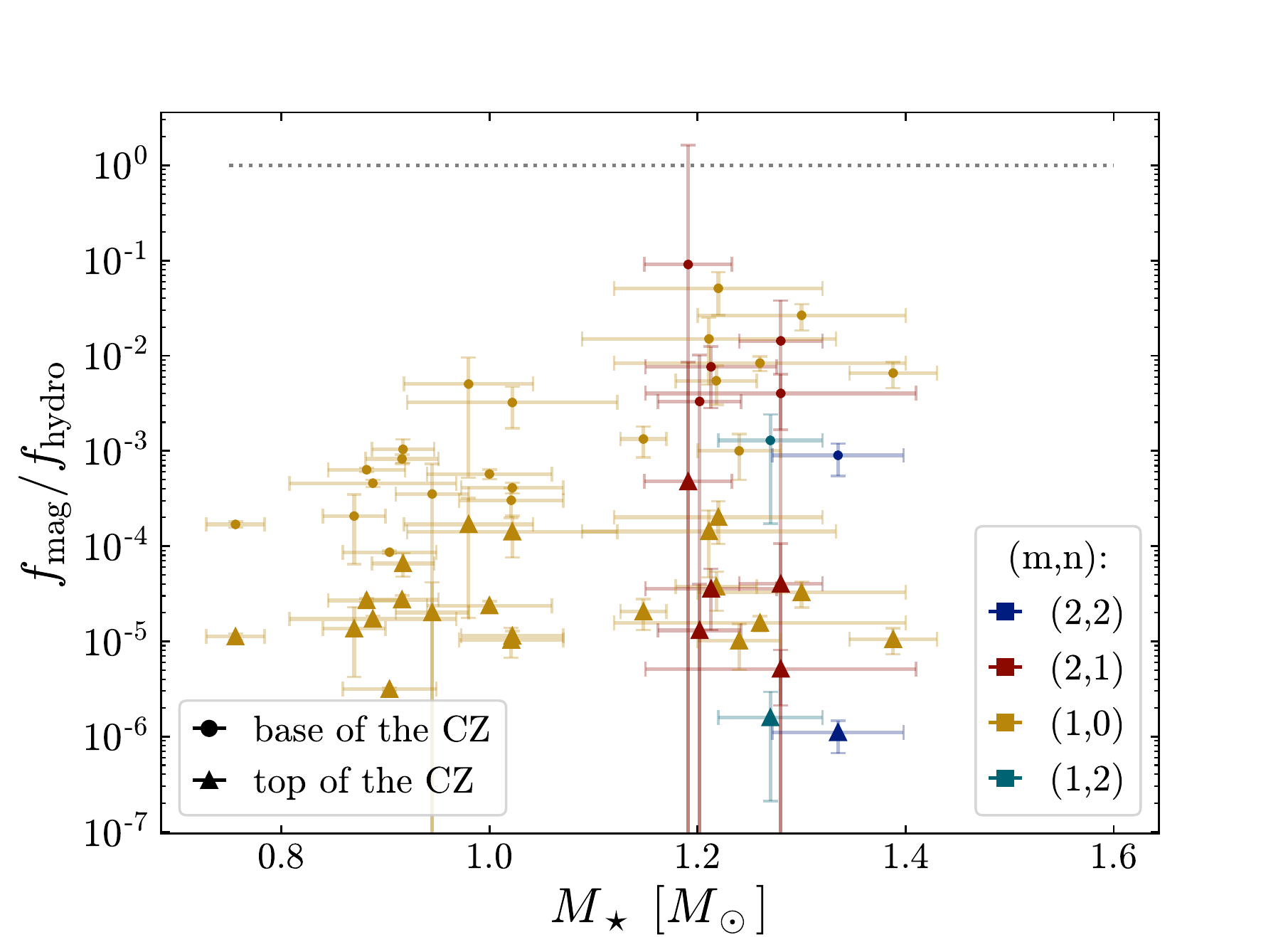}
\caption{Dependence of the ratio of the Lorentz forcing relative to the hydrodynamic forcing with respect to  the mass of the host star for the selected star-planet systems (see Table \ref{tabsys}).
The magnetic forcing term $\fm$  is estimated with the magnetostrophic balance at the base and the top of the CZ and a stellar model evolved with  a slow initial rotation. The tidal frequency has been calculated with the pair $(m,n)$ that minimizes $\Roh\scm$ and thus maximizes $\fm/\fh$. 
}
\label{fVSm}
\end{figure}

In Fig. \ref{fVSm}, we present the ratio of the magnetic to hydrodynamic forcings $\fm/\fh=\Le^2\Roh^{-1}\scm^{-1}$ as a  function of the mass of the host star. The quantity $\Roh\scm$ has been calculated using orbital and stellar rotation periods  (see Table \ref{tabsys}) for the pair $(m,n)$ that minimizes this quantity and thus maximizes $\fm/\fh$. To calculate the Lehnert number squared, we have used the star's rotation period and radius displayed in Table \ref{tabsys}, coupled with density, convective length and velocity given by our grid of STAREVOL models at the closest age and mass of the host star. For each system, $\Le^2$ is evaluated at the base and the top of the CZ. 
Moreover,  the magnetostrophic regime has been selected because it best reproduces the observed surface magnetic fields (see Figs. \ref{fig1} and \ref{BobsBdyn}).  
As expected, the ratio $\fm/\fh$ at the base is greater than at the top of the CZ, consistent with the relative magnitude of the  large-scale magnetic field inside the convective envelope. Moreover, at the base of the CZ the higher the mass, the greater this ratio. On the contrary, we observe a drop in $\fm/\fh$ at the top of the CZ for stars more massive than $1.2\ M_\odot$, similar to what we notice in Fig. \ref{fig3}.
More importantly, we point out that $\fm/\fh$ is always smaller than unity, regardless of the mass of star and the position inside the stellar envelope. 
Only HAT-P-24 (b) may feature  the ratio of forcings around unity within the error bar  for $(m,n)=(2,1)$ at the base of the CZ. Indeed, the rotation period of the star HAT-P-24 is nearly  twice the orbital period of the planet HAT-P-24 b, which implies that $\Roh$ is very close to zero, thereby making $\fm/\fh$ close to unity, despite the small value of $\Le^2$. 
 It is worth noticing  that \cite{O2009} and \cite{LO2018} mentioned a significant impact of the magnetic field on non-wave like motions as soon as $\Le>0.1$. In our study, some star-planet systems (like HAT-P-13 and HD 149026) feature a Lehnert number greater than $10^{-1}$ and yet a ratio $\fm/\fh$ far below unity.  This stresses the fact that the Lehnert number is not the only quantity to come into play and that the relative amplitude of the tidal and rotation frequencies has to be considered to conclude on the impact of magnetism on the tidal forcing.\\

From the results of this section, we conclude that the tidal forcing arising from the Lorentz force remains small in comparison to a pure hydrodynamical forcing. This conclusion is important, as it stresses that adopting a Coriolis-driven tidal forcing is justified to study the propagation and dissipation of tidally-forced magneto-inertial waves in the convective envelope of low-mass stars, despite the presence of a large-scale, dynamo generated magnetic field, as was done in  \cite{W2016,W2018} and  \cite{LO2018}.
\section{The relative importance of viscous over Ohmic  dissipation for  magneto-inertial waves}
\label{sec3}
In the previous section, we have demonstrated that the Lorentz force has a weak contribution to the tidal forcing of (magneto-) inertial waves. It can,  however, affect their propagation (hence the name magneto-inertial waves)  and dissipation  as the Lorentz force acts on the wave-like part of the equation of motion and we have introduced the Ohmic diffusion in the induction equation (see Eqs. (\ref{de_eq}) and (\ref{eq01b}), respectively).  \cite{W2016} studied the dissipation of these waves by turbulent friction and magnetic diffusion processes  using a local Cartesian model of an isentropic  convective region. He  compared the importance of Ohmic versus  viscous dissipations, especially at resonances. 
Varying the Lehnert number, \cite{W2016} found that the transition from a viscous-dominated regime to a regime dominated by Ohmic dissipation occurs when the Lehnert number is greater than  $\sim10^{-4}$--$10^{-3}$, for average atomic Ekman and Prandtl numbers close to those  expected in the Sun or in Jupiter. This led him to the conclusion that when $\Le$ is larger than $10^{-3}$, magnetic effects on tidal dissipation should be taken into account.

This work has been taken up by \cite{LO2018} in which they studied the propagation and the kinetic and magnetic energy dissipations of tidally forced magneto-inertial waves in a  spherical shell.
They showed that, at high Lehnert numbers, dissipation is no longer  focused along the  shear layers that are  shaped by rotation and viscosity following attractors of characteristics as  in the pure hydrodynamical case.
\begin{figure}[h]
\centering
\resizebox{4\hsize/5}{!}{\includegraphics[scale=0.5]{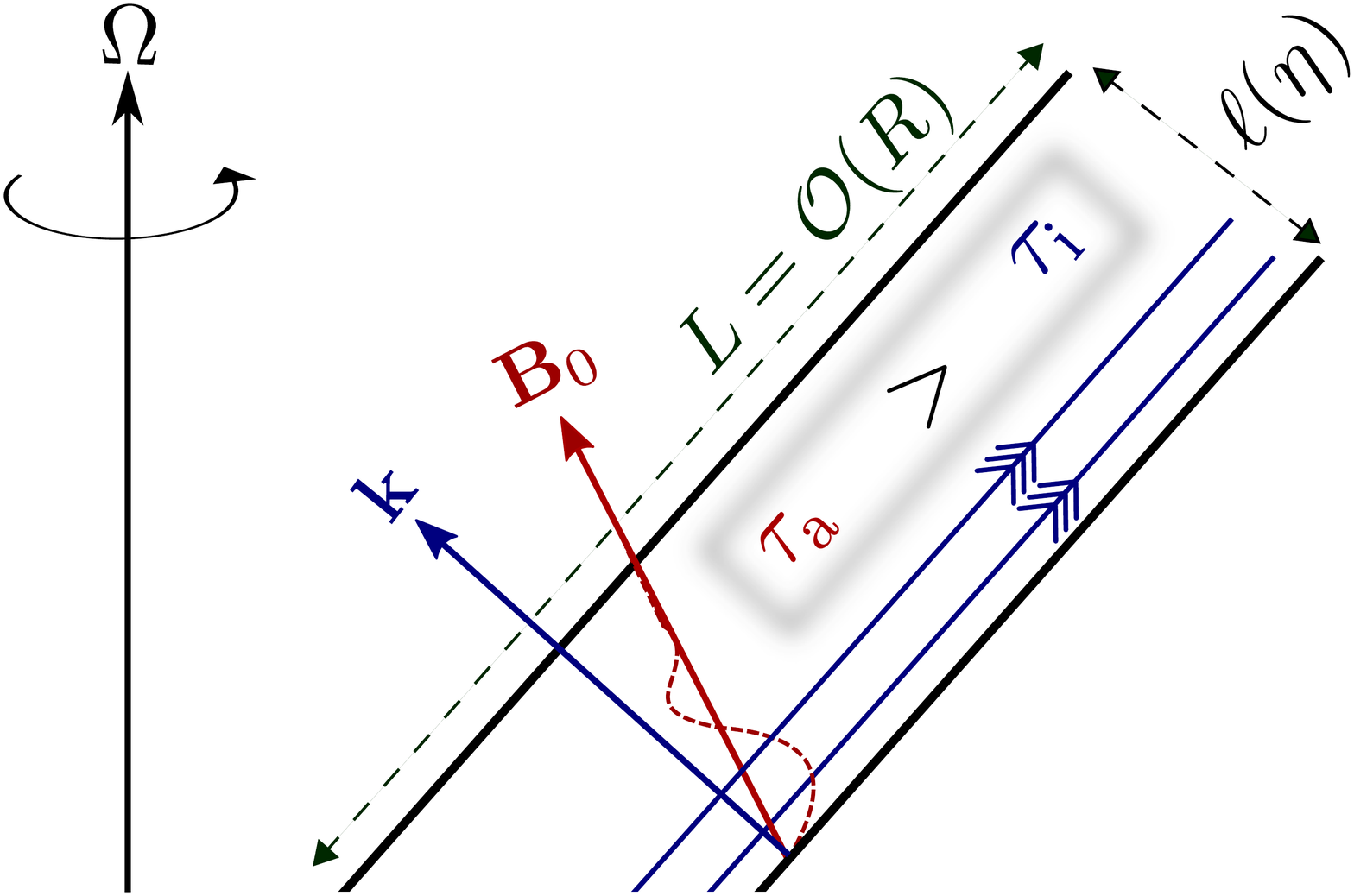}}
\caption{Sketch of an inertial wave  beam with a magnetic field (\textit{red arrow}), an Alfvén wave (\textit{dashed red line}) and an inertial wave propagating  from the left to the right (\textit{blue arrows)}. The time for an Alfvén wave to transversely cross the wave  beam of length $\ell(\tau_\mathrm{i}=\tau_\eta)$, is greater than the time for an inertial wave to go through the wave  beam of length $L\sim R$.}
\label{fig5}
\end{figure}
\begin{figure*}[ht]
\resizebox{\hsize/2}{!}{\includegraphics{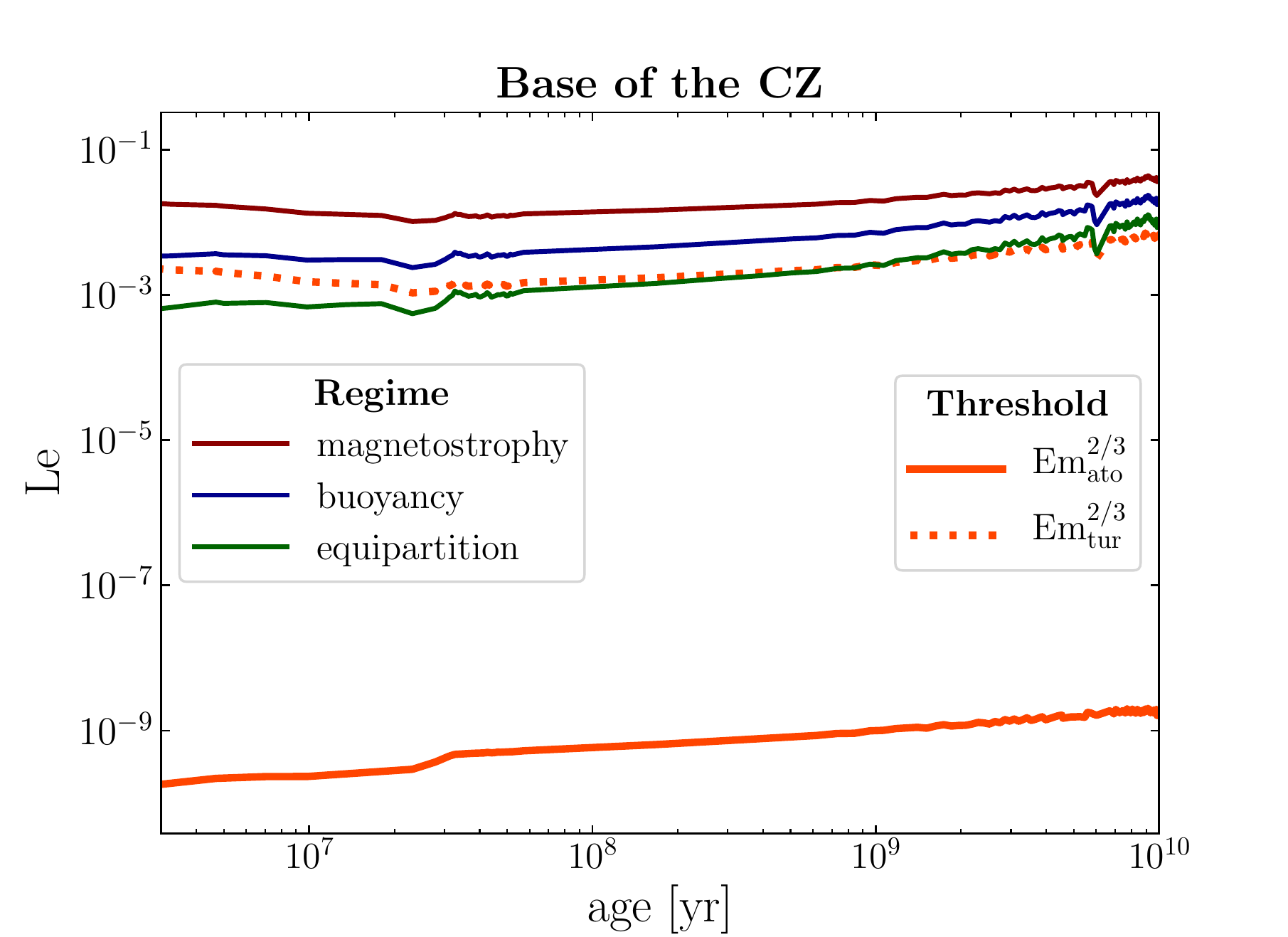}}
\resizebox{\hsize/2}{!}{\includegraphics{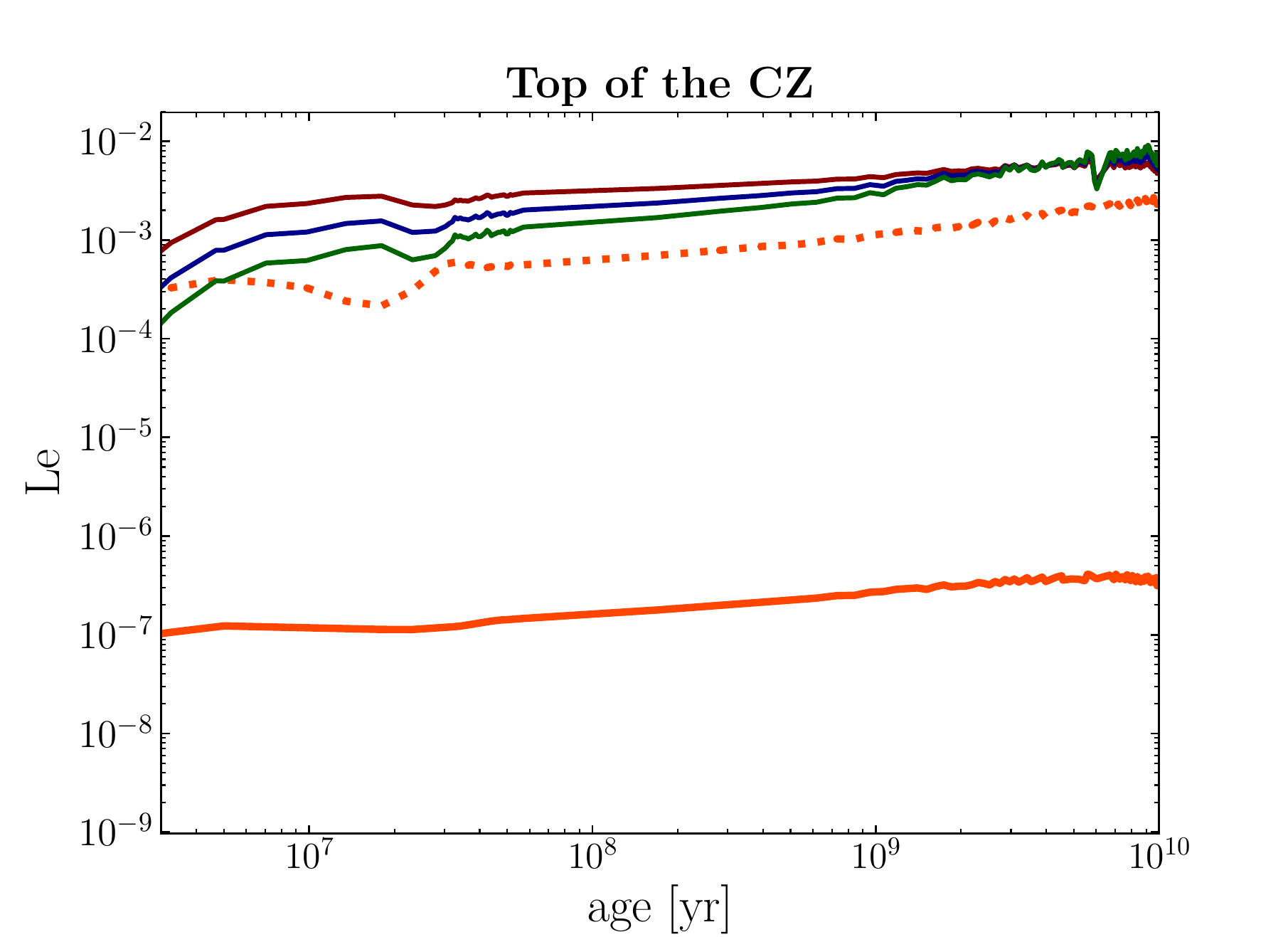}}
\caption{Lehnert number at the base (\textit{left panel}) and top (\textit{right panel}) of the convective zone  versus the age of a $1M_\odot$ star. Slow initial rotation has been used in both panels. The thresholds for which magnetic field impacts the propagation of inertial waves  is estimated with different Ekman numbers  and drawn in orange (see the legend).}
\label{fig4}
\end{figure*}
Once again, they identified a critical Lehnert number that separates the  regimes dominated by viscosity or by Ohmic dissipation. They expressed this critical Lehnert number with the help of characteristic timescales of ray tracing. Specifically, the width of an inertial wave  beam $\ell$ is deduced by equating the magnetic diffusion timescale $\tau_\eta=\ell^2/\eta$ and the  inertial wave propagation time $\tau_\mathrm{i}=R/|{\bf V}_\mathrm{g}|$, where $|{\bf V}_\mathrm{g}|\sim\ell\Omega$ is the group velocity (see Fig. \ref{fig5}). Indeed, the higher the Ohmic diffusion (or the viscosity), the larger  the spread of the inertial wave beam. Furthermore, hydrodynamical terms  prevail over MHD ones  when Alfvén waves (produced by the deformation of the magnetic field by the  inertial flow) do not have time to distort the wave  beam. In other words the hydrodynamical terms  dominate when
$\tau_\mathrm{a}>\tau_\mathrm{i}$ \text{ where } 
\[
\tau_\mathrm{a}=\ell/|{\bf V}_\mathrm{a}|\sim\Le^{-1}\Omega^{-1}\ell/R,
\]
 is the typical time for an Alfvén wave to transversely cross the wave  beam (see Fig. \ref{fig5}).
Using these heuristic considerations, \cite{LO2018} have shown that the propagation of inertial waves  is little influenced by a magnetic field  as long as 
\[\Le\ll\Em^{2/3},\]
 with  $\Em=\eta/(2\Omega R^2)$ the magnetic Ekman number.
This prediction has been inferred in the context of a low $\mathrm{Pm}$. This condition is generally  satisfied  in solar-like stars. In the Sun, the  atomic   magnetic Prandtl number varies  from $10^{-6}$ at the surface to $10^{-1}$ at the  base of the CZ \citep{Z1983}. 
When viscosity dominates, for instance in the core of massive stars \citep[see Fig. 2 in ][]{AB2019}, the same relationship holds with  the viscous Ekman number $\Ek$ instead of $\Em$.

In Fig. \ref{fig4}, the Lehnert number is illustrated as a function of stellar age, where the threshold defined in \cite{LO2018} is included with atomic and turbulent magnetic diffusivities:  $\Em_\mathrm{ato}^{2/3}$ and $\Em_\mathrm{tur}^{2/3}$ being the atomic and turbulent magnetic Ekman numbers, respectively. 
 The parameter $\Em_\mathrm{ato}$ has been computed thanks to the Braginskii prescription for plasma diffusivities \citep[Appendix B]{B1965,AB2019} using the grid of STAREVOL models. 
The turbulent magnetic Ekman number ($\Em_\mathrm{tur}$) is derived assuming that the eddy-magnetic  diffusivity takes the simple form: $\eta_\mathrm{tur}=\uc\lc/3$. In this approach, $\eta_\mathrm{tur}$ is equivalent to the eddy-viscosity that means a magnetic turbulent Prandtl number of the order of unity \citep[see e.g.][]{CT1992}. 
We highlight that in both panels of Fig. \ref{fig4}, the  threshold calculated with an atomic magnetic diffusivity is much lower than $\Le$ derived in the various regimes, by at least three order of magnitudes near the surface and six order of magnitude at the base of the CZ. 
Consequently, the Ohmic dissipation largely outbalances the viscous dissipation along the lifetime of a 1$M_\odot$ when taking Lin \& Ogilvie's threshold assessed with a magnetic atomic diffusivity $\eta_\mathrm{ato}$.
In contrast, the limit estimated with the turbulent magnetic Ekman number is of the same order of magnitude as the Lehnert number in the equipartition regime at the base of the CZ, whereas the threshold is slightly smaller than the Lehnert number curves at the top. Note that the turbulent magnetic Ekman number is close to the value of the magnetic Ekman number used in Lin \& Ogilvie's (previously quoted) paper for their  simulations. This explains why the transition from an hydrodynamical to a fully magnetic regime is carried out at similar Lehnert number in their case and in ours when using $\Em_\mathrm{tur}$ here \citep[see Figs. 2, 3 and 4 in ][]{LO2018}.  
By choosing an eddy-magnetic diffusivity, both Ohmic and viscous dissipations have to be taken into account in the dissipation calculation.
\section{The impact of a smaller-scale magnetic field}
\label{sec4}

The stellar dynamo is a multi-scale process. Indeed, large-scale and small-scale dynamos coexist inside a star's  CZ and produce magnetic fields at different length scales \cite[and references therein]{BS2005,SB2013}.
  At the base of the CZ, the convective length scale computed by STAREVOL  is about one tenth of the radius of the star and decreases drastically towards the top of the CZ ($\sim10^{-4}\,R$ in our model).  In Sect. \ref{sec22}, we have made the assumption that a dynamo-like magnetic field is the result of turbulent convective motions featured by the convective velocity at the base of the CZ, associated to a relatively large-scale convective length. Then, we have used this dynamo-induced magnetic field to extrapolate a dipolar magnetic field near the surface of the star (see Eq. (\ref{Bdip})). 
However, one can question the role of the small-scale dynamo fields on tidal excitation and dissipation  throughout the convective envelope. 

\begin{figure}[ht]
\resizebox{\hsize}{!}{\includegraphics[scale=0.5]{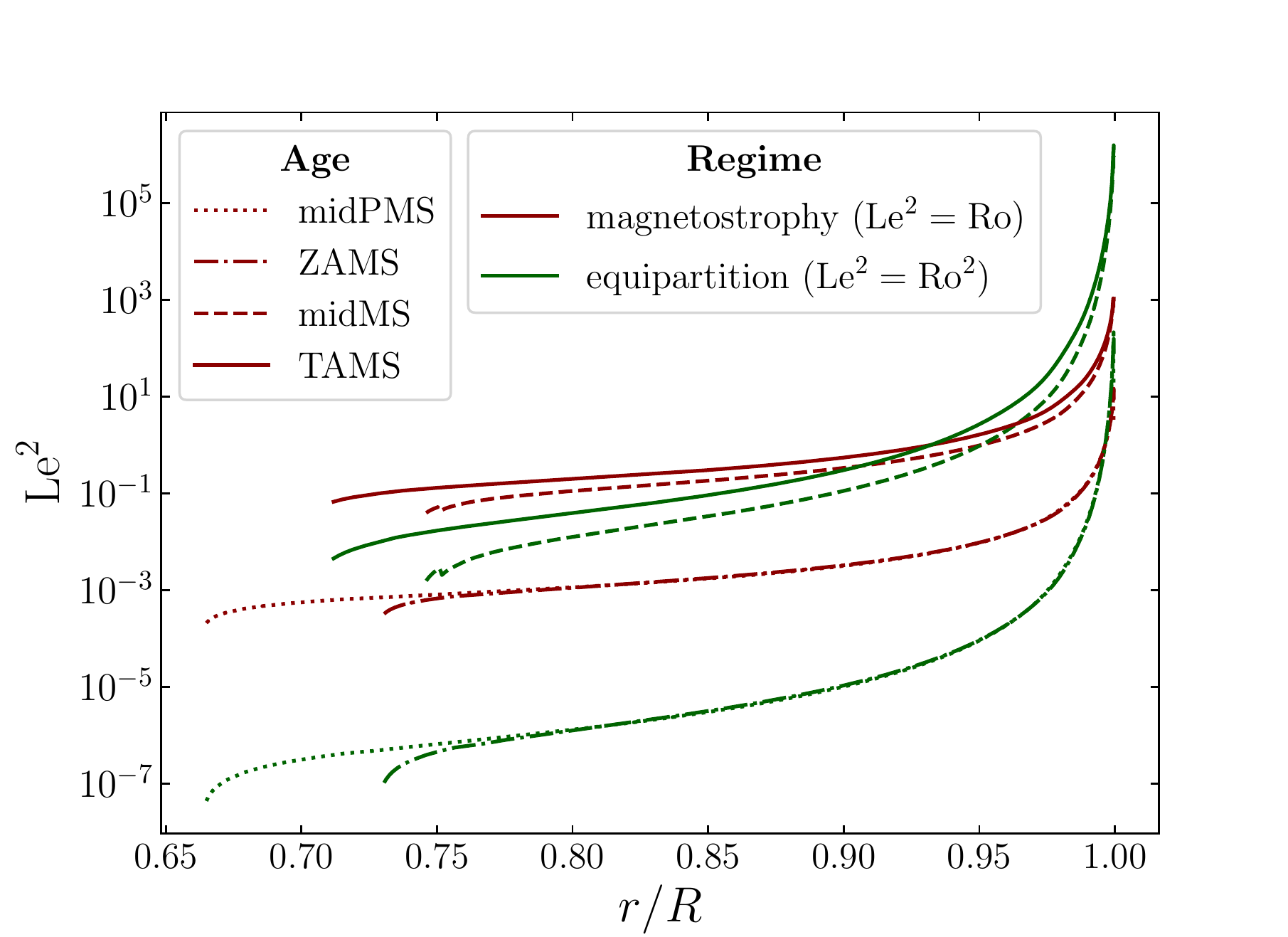}}
\caption{Lehnert number squared against the normalised radius in the magnetostrophic (\textit{dark red}) and equipartition (\textit{green}) regimes,  for a $1M_\odot$ star. \bfo The reader is referred to the body of Sect. \ref{sec4} for the new definition of $\Le$ that involves a small-scale magnetic field. 
The Lehnert number squared \bfc is plotted  at  different epochs: amid the pre-main sequence (midPMS; $\sim30\,\mathrm{Myr}$), at zero-age main sequence (ZAMS; $\sim60\,\mathrm{Myr}$), amid the main sequence (midMS; $\sim5.3\,\mathrm{Gyr}$), and toward the end of the main sequence (TAMS; $\sim10.5\,\mathrm{Gyr}$).}
\label{Le2_r}
\end{figure}

\bfo The dynamical tides will possibly interact with the smaller-scale magnetic field. Indeed, the scale of variation of dynamical tides along the inertial wave beam is of order $R$ (Sect. \ref{sec3}), but the transverse length scale $l$ of this beam is smaller. The balance between magnetic diffusion and inertial timescales (see also Sect. \ref{sec3}) leads to $l=\sqrt[3]\Em\,R$ i.e. the transverse length scale of the beam varies from one hundredth to one tenth of the stellar radius  when using typical values of the turbulent Ekman number (see Fig. \ref{fig4}). This orthogonal length scale is thus comparable to the length scale of the convection and of the corresponding magnetic field.

The scale of variation of the equilibrium tide is also of order $R$ \citep[see e.g.][]{RM2012}. The small-scale components of the star's magnetic field can collectively affect the large-scale flow of the equilibrium tide provided their correlations sustain a large-scale effective Lorentz force. \bfc
We plot $\Le^2$ associated with \bfo a \bfc small-scale magnetic field in Fig. \ref{Le2_r} as a function of the normalized radius in the whole CZ. \bfo The typical length to be used in the expression of $\fm$ (Eq. (\ref{eq1})) is no longer $R$ but $l_\mathrm{c}$ which represents the characteristic length of the fluctuating component of the dynamo-generated magnetic field. 
We redefine the Lehnert number here as $\Le=B_\mathrm{dyn}/(\sqrt{\mu_0\rho}2\Omega\lc)$, where $B_\mathrm{dyn}$ (Table \ref{tab1}) depends upon $\lc$, which in turn varies inside the CZ. In the three regimes listed in Table \ref{tab2}, $\Le$ now depends only on the Rossby number (i.e. each scaling has to be multiplied by $R/\lc$). \bfc
 The profile of $\Le^2$ is shown for magnetostrophic and equipartition regimes at different evolutionary stages. In \bfo both regimes\bfc, the Lehnert number follows the Rossby number tendency obtained in \cite{MA2016}. Indeed, the authors  pointed out that $\Ro$ always increases with radius, regardless of the changes in mass and stellar evolutionary phases. \bfo In the early stages of the $1M_\odot$ star's evolution (PMS and ZAMS), the probability of having a significant magnetic forcing e.g. $\fm/\fh>1$ is weak at the base of the CZ ($\Le^2<10^{-3}$),  but becomes high near the surface ($\Le^2>10$) based on the analyses of Sects. \ref{sec25} and \ref{sec26}. The chances are even higher when the star is older (from the midpoint to the end of the MS)  since $\Le^2$ is already greater than $10^{-3}$ at the base of the CZ in both regimes.
\bfc

From this analysis, we should  keep in mind that \bfo small-scale magnetic fields can affect the equilibrium and dynamical tides. \bfc Indeed, $\Le^2$ can be significantly enhanced near the surface of the star, \bfo a fortiori for all three regimes described in Table \ref{tab1}\bfc. This may change the results of Sect. \ref{sec26}, namely boost the effect of magnetism on tidal forcing so that $\fm$ is no longer negligible in front of $\fh$. 

\section{Conclusions and discussions}
\label{sec_conclu}
The influence of magnetism on tidal interaction along the evolution of low-mass stars  has been investigated through its impact on tidal excitation and dissipation. We have first derived an analytic criterion to quantify the Lorentz contribution to the  tidal forcing. The relative importance of Coriolis-like versus Lorentz-like forcings relies on the Lehnert number and characteristic frequencies of star-planet systems, i.e. the tidal frequency and the rotation frequency of the main body.
With the help of simple scaling laws, we have estimated the magnitude of a dynamo-generated  magnetic field near the tachocline. Then, a large-scale dipolar magnetic field has been inferred near the surface of the star from these scaling relationships.  
For this purpose, we have used grids of low-mass stars computed by the stellar evolution code STAREVOL \citep{AP2019}. This allowed us to compare our magnetic field estimates at the surface with the observations of the surface dipolar magnetic field. We find  that the  magnetostrophic regime (resulting from the balance between  Coriolis and Lorentz forces)  gives a good estimate, within an order of magnitude, of the magnetic fields of PMS and MS stars  when assuming a median or fast initial rotation. Nevertheless, 
improvements still need  to be made in terms of magnetic scaling laws  in order to better reproduce  the observed magnetic fields of \cite{VG2014} and \cite{SJ2017} (see Appendix \ref{dis}).

Subsequently, we have estimated the Lehnert number with these  prescriptions for  the stellar dynamo at the base and the top of the convective zone. We have explored how it varies for different stellar masses and initial rotation speeds  as a function of  time. In all cases, the Lehnert number is small compared to  unity. This means  that magnetism can play a role in tidal forcing only for small tidal to rotational frequency ratios. 
  In practice, this condition  is verified neither at the base nor at the top of the convective zone when  using the  large-scale magnetic fields in the primary component of  quasi-circular and coplanar star-planet systems. Although the ratio $\fm/\fh$ increases with the mass of the host star until $1.2M_\odot$ (see Fig. \ref{fVSm}) and is greater at the base of the convective zone, it remains below $1$ for all the studied systems.
Indeed, only specific and sharp enough resonances between the tidal and the rotational frequency can increase the $\fm/\fh$ ratio. This statement is also valid for non-circular and non-coplanar systems, the tidal frequency being just a different combination of orbital and rotation frequencies. Eventually,  one should note that all the selected systems (Table \ref{tabsys}) have a host star on the mid and late MS. But it should not change the fact that large-scale magnetic fields have little influence on the tidal excitation because $\Le^2$ is even  weaker for younger  stars (as we have seen in Fig. \ref{fig3}).

In contrast, the dissipation of the dynamical tide is strongly affected by stellar  magnetism. Indeed,  we have shown that viscous dissipation is no longer the main process of energy dissipation. Ohmic dissipation is at least as important as viscous dissipation (see Fig. \ref{fig4}) in the whole  convective zone for a $1M_\odot$ star as well as other low-mass stars.
This means that a full MHD treatment is needed to analyse the propagation and dissipation of tidally-forced (magneto-) inertial waves inside the convective zone of a low-mass star.

It can be added that whatever the energy mix distributed in the toroidal and poloidal components of the large-scale magnetic field at the surface \citep[which is also variable over time  ][]{KP2017}, it does not change the main conclusions of our paper because these energies are comparable within an order of magnitude \citep{SJ2015}.\\



In Sect. \ref{sec4}, we have addressed the question of the effect of \bfo small-scale components of the dynamo-induced magnetic field on tidal interactions. We have pointed out that they are likely to play a role on both dynamical and equilibrium tides and consequently on the tidal forcing and dissipation. \bfc  
\bfo We first demonstrated that the transverse component of \bfc the typical length scale of the \bfo dynamical tide is commensurate with \bfc the convective  mixing length in the \bfo bulk of the \bfc convective zone.
\bfo In this respect, it would \bfc also mean that the modelling of the  friction applied by turbulent convection on the \bfo dynamical tides \bfc should go beyond  an eddy viscosity model \citep{OL2012} \bfo to incorporate  magnetic effects. 
Then, we showed that the Lehnert number squared estimated \bfo with a small-scale magnetic field \bfc can be much greater than $\Le^2$ calculated with a large-scale field. As a result, the impact of a small-scale dynamo induced magnetic field on tidal forcing could be not as negligible as the effect of a larger-scale magnetic field. Quantifying this impact now requires ab-initio modelling of tidal forcing in turbulent convective envelopes sustaining a small-scale dynamo.


\begin{acknowledgements}
\bfo We would like to thank the anonymous referee for the helpful comments and suggestions regarding our work. \bfc A. Astoul, K. Augustson,  E. Bolmont,  and S. Mathis acknowledge funding by the European Research Council through the ERC grant SPIRE 647383.
The authors acknowledge the PLATO CNES funding at CEA/IRFU/DAp and IRAP. The authors further thank V. See for fruitful discussions and the use of his data. 
F. Gallet acknowledges financial support from a  CNES fellowship. A.S.Brun acknowledges funding by ERC WHOLESUN 810218 grant, INSU/PNST, and CNES Solar Orbiter. 
This work has been carried out within the framework of the NCCR PlanetS supported by the Swiss National Science Foundation. This research has made use of NASA's Astrophysics Data System. 
\end{acknowledgements}

\bibliographystyle{aa.bst} 
\bibliography{biblio.bib}

\begin{thebibliography}{112}
\expandafter\ifx\csname natexlab\endcsname\relax\def\natexlab#1{#1}\fi

\bibitem[{{Amard} {et~al.}(2016){Amard}, {Palacios}, {Charbonnel}, {Gallet}, \&
  {Bouvier}}]{AP2016}
{Amard}, L., {Palacios}, A., {Charbonnel}, C., {Gallet}, F., \& {Bouvier}, J.
  2016, \aap, 587, A105

\bibitem[{{Amard} {et~al.}(2019){Amard}, {Palacios}, {Charbonnel}, {Gallet},
  {Georgy}, {Lagarde}, \& {Siess}}]{AP2019}
{Amard}, L., {Palacios}, A., {Charbonnel}, C., {et~al.} 2019, arXiv e-prints,
  arXiv:1905.08516

\bibitem[{{Anderson} {et~al.}(2011){Anderson}, {Barros}, {Boisse}, {Bouchy},
  {Collier Cameron}, {Faedi}, {Hebrard}, {Hellier}, {Lendl}, {Moutou},
  {Pollacco}, {Santerne}, {Smalley}, {Smith}, {Todd}, {Triaud}, {West},
  {Wheatley}, {Bento}, {Enoch}, {Gillon}, {Maxted}, {McCormac}, {Queloz},
  {Simpson}, \& {Skillen}}]{AB2011}
{Anderson}, D.~R., {Barros}, S.~C.~C., {Boisse}, I., {et~al.} 2011,
  Publications of the Astronomical Society of the Pacific, 123, 555

\bibitem[{{Anderson} {et~al.}(2014){Anderson}, {Brown}, {Collier Cameron},
  {Delrez}, {Fumel}, {Gillon}, {Hellier}, {Jehin}, {Lendl}, {Maxted},
  {Neveu-VanMalle}, {Pepe}, {Pollacco}, {Queloz}, {Rojo}, {Segransan},
  {Serenelli}, {Smalley}, {Smith}, {Southworth}, {Triaud}, {Turner}, {Udry}, \&
  {West}}]{AB2014}
{Anderson}, D.~R., {Brown}, D.~J.~A., {Collier Cameron}, A., {et~al.} 2014,
  arXiv e-prints, arXiv:1410.3449

\bibitem[{{Antia} {et~al.}(2000){Antia}, {Chitre}, \& {Thompson}}]{A2000}
{Antia}, H.~M., {Chitre}, S.~M., \& {Thompson}, M.~J. 2000, \aap, 360, 335

\bibitem[{{Asplund} {et~al.}(2009){Asplund}, {Grevesse}, {Sauval}, \&
  {Scott}}]{AG2009}
{Asplund}, M., {Grevesse}, N., {Sauval}, A.~J., \& {Scott}, P. 2009, Annual
  Review of Astronomy and Astrophysics, 47, 481

\bibitem[{{Augustson} {et~al.}(2019){Augustson}, {Brun}, \& {Toomre}}]{AB2019}
{Augustson}, K.~C., {Brun}, A.~S., \& {Toomre}, J. 2019, \apj, 876, 83

\bibitem[{{Barker} \& {Ogilvie}(2010)}]{BO2010}
{Barker}, A.~J. \& {Ogilvie}, G.~I. 2010, \mnras, 404, 1849

\bibitem[{{Basu}(2016)}]{B2016}
{Basu}, S. 2016, Living Reviews in Solar Physics, 13, 2

\bibitem[{{Beck} {et~al.}(2018){Beck}, {Mathis}, {Gallet}, {Charbonnel},
  {Benbakoura}, {Garc{\'\i}a}, \& {do Nascimento}}]{BM2018}
{Beck}, P.~G., {Mathis}, S., {Gallet}, F., {et~al.} 2018, \mnras, 479, L123

\bibitem[{{Bell} {et~al.}(2013){Bell}, {Naylor}, {Mayne}, {Jeffries}, \&
  {Littlefair}}]{BN2013}
{Bell}, C. P.~M., {Naylor}, T., {Mayne}, N.~J., {Jeffries}, R.~D., \&
  {Littlefair}, S.~P. 2013, \mnras, 434, 806

\bibitem[{{Benbakoura} {et~al.}(2019){Benbakoura}, {R{\'e}ville}, {Brun}, {Le
  Poncin-Lafitte}, \& {Mathis}}]{BR2019}
{Benbakoura}, M., {R{\'e}ville}, V., {Brun}, A.~S., {Le Poncin-Lafitte}, C., \&
  {Mathis}, S. 2019, \aap, 621, A124

\bibitem[{{Bolmont} {et~al.}(2017){Bolmont}, {Gallet}, {Mathis}, {Charbonnel},
  {Amard}, \& {Alibert}}]{BG2017}
{Bolmont}, E., {Gallet}, F., {Mathis}, S., {et~al.} 2017, \aap, 604, A113

\bibitem[{{Bolmont} \& {Mathis}(2016)}]{BM2016}
{Bolmont}, E. \& {Mathis}, S. 2016, Celestial Mechanics and Dynamical
  Astronomy, 126, 275

\bibitem[{{Braginskii}(1965)}]{B1965}
{Braginskii}, S.~I. 1965, Reviews of Plasma Physics, 1, 205

\bibitem[{{Brandenburg} \& {Subramanian}(2005)}]{BS2005}
{Brandenburg}, A. \& {Subramanian}, K. 2005, \physrep, 417, 1

\bibitem[{{Brown}(2014)}]{B2014}
{Brown}, D.~J.~A. 2014, \mnras, 442, 1844

\bibitem[{{Brun} \& {Browning}(2017)}]{BB2017}
{Brun}, A.~S. \& {Browning}, M.~K. 2017, Living Reviews in Solar Physics, 14, 4

\bibitem[{{Brun} {et~al.}(2015){Brun}, {Garc{\'{\i}}a}, {Houdek}, {Nandy}, \&
  {Pinsonneault}}]{BG2015}
{Brun}, A.~S., {Garc{\'{\i}}a}, R.~A., {Houdek}, G., {Nandy}, D., \&
  {Pinsonneault}, M. 2015, \ssr, 196, 303

\bibitem[{{Camargo} \& {Tasso}(1992)}]{CT1992}
{Camargo}, S.~J. \& {Tasso}, H. 1992, Physics of Fluids B, 4, 1199

\bibitem[{{Charbonneau}(2013)}]{C2013}
{Charbonneau}, P. 2013, Solar and Stellar Dynamos: Saas-Fee Advanced Course 39
  Swiss Society for Astrophysics and Astronomy, Saas-Fee Advanced Courses,
  Volume 39.~ISBN 978-3-642-32092-7.~Springer-Verlag Berlin Heidelberg, 2013,
  39

\bibitem[{{Charbonneau}(2014)}]{C2014}
{Charbonneau}, P. 2014, \araa, 52, 251

\bibitem[{{Christensen} {et~al.}(2009){Christensen}, {Holzwarth}, \&
  {Reiners}}]{CH2009}
{Christensen}, U.~R., {Holzwarth}, V., \& {Reiners}, A. 2009, \nat, 457, 167

\bibitem[{{Cowling}(1941)}]{C1941}
{Cowling}, T.~G. 1941, \mnras, 101, 367

\bibitem[{{Czesla} {et~al.}(2017){Czesla}, {Salz}, {Schneider}, {Mittag}, \&
  {Schmitt}}]{CS2017}
{Czesla}, S., {Salz}, M., {Schneider}, P.~C., {Mittag}, M., \& {Schmitt},
  J.~H.~M.~M. 2017, \aap, 607, A101

\bibitem[{{Davidson}(2013)}]{D2013}
{Davidson}, P.~A. 2013, Geophysical Journal International, 195, 67

\bibitem[{DeRosa {et~al.}(2012)DeRosa, Brun, \& Hoeksema}]{DB2012}
DeRosa, M.~L., Brun, A.~S., \& Hoeksema, J.~T. 2012, The Astrophysical Journal,
  757, 96

\bibitem[{{Donati} {et~al.}(2006){Donati}, {Catala}, {Landstreet}, \&
  {Petit}}]{D2006}
{Donati}, J.-F., {Catala}, C., {Landstreet}, J.~D., \& {Petit}, P. 2006, in
  Astronomical Society of the Pacific Conference Series, Vol. 358, Astronomical
  Society of the Pacific Conference Series, ed. R.~{Casini} \& B.~W. {Lites},
  362

\bibitem[{{Donati} {et~al.}(2007){Donati}, {Jardine}, {Gregory}, {Petit},
  {Bouvier}, {Dougados}, {M{\'e}nard}, {Collier Cameron}, {Harries}, {Jeffers},
  \& {Paletou}}]{DJ2007}
{Donati}, J.-F., {Jardine}, M.~M., {Gregory}, S.~G., {et~al.} 2007, \mnras,
  380, 1297

\bibitem[{{Donati} \& {Landstreet}(2009)}]{DL2009}
{Donati}, J.-F. \& {Landstreet}, J.~D. 2009, \araa, 47, 333

\bibitem[{{Duez} {et~al.}(2010){Duez}, {Mathis}, \&
  {Turck-Chi{\`e}ze}}]{DM2010}
{Duez}, V., {Mathis}, S., \& {Turck-Chi{\`e}ze}, S. 2010, \mnras, 402, 271

\bibitem[{{Emeriau-Viard} \& {Brun}(2017)}]{EB2017}
{Emeriau-Viard}, C. \& {Brun}, A.~S. 2017, \apj, 846, 8

\bibitem[{{Esposito} {et~al.}(2017){Esposito}, {Covino}, {Desidera}, {Mancini},
  {Nascimbeni}, {Zanmar Sanchez}, {Biazzo}, {Lanza}, {Leto}, {Southworth},
  {Bonomo}, {Su{\'a}rez Mascare{\~n}o}, {Boccato}, {Cosentino}, {Claudi},
  {Gratton}, {Maggio}, {Micela}, {Molinari}, {Pagano}, {Piotto}, {Poretti},
  {Smareglia}, {Sozzetti}, {Affer}, {Anderson}, {Andreuzzi}, {Benatti},
  {Bignamini}, {Borsa}, {Borsato}, {Ciceri}, {Damasso}, {di Fabrizio},
  {Giacobbe}, {Granata}, {Harutyunyan}, {Henning}, {Malavolta}, {Maldonado},
  {Martinez Fiorenzano}, {Masiero}, {Molaro}, {Molinaro}, {Pedani}, {Rainer},
  {Scandariato}, \& {Turner}}]{EC2017}
{Esposito}, M., {Covino}, E., {Desidera}, S., {et~al.} 2017, \aap, 601, A53

\bibitem[{{Ferraz-Mello} {et~al.}(2015){Ferraz-Mello}, {Tadeu dos Santos},
  {Folonier}, {Czismadia}, {do Nascimento}, \& {P{\"a}tzold}}]{FT2015}
{Ferraz-Mello}, S., {Tadeu dos Santos}, M., {Folonier}, H., {et~al.} 2015,
  \apj, 807, 78

\bibitem[{Finlay(2008)}]{F2008}
Finlay, C.~C. 2008, in Les Houches, Vol.~88, Dynamos, ed. P.~Cardin \&
  L.~Cugliandolo (Elsevier), 403 -- 450

\bibitem[{{Folsom} {et~al.}(2016){Folsom}, {Petit}, {Bouvier}, {L{\`e}bre},
  {Amard}, {Palacios}, {Morin}, {Donati}, {Jeffers}, \& {Marsden}}]{FP2016}
{Folsom}, C.~P., {Petit}, P., {Bouvier}, J., {et~al.} 2016, \mnras, 457, 580

\bibitem[{{Gallet} {et~al.}(2017){Gallet}, {Bolmont}, {Mathis}, {Charbonnel},
  \& {Amard}}]{G2017}
{Gallet}, F., {Bolmont}, E., {Mathis}, S., {Charbonnel}, C., \& {Amard}, L.
  2017, \aap, 604, A112

\bibitem[{{Gallet} \& {Bouvier}(2013)}]{GB2013}
{Gallet}, F. \& {Bouvier}, J. 2013, \aap, 556, A36

\bibitem[{{Gallet} \& {Bouvier}(2015)}]{GB2015}
{Gallet}, F. \& {Bouvier}, J. 2015, \aap, 577, A98

\bibitem[{{Gandolfi} {et~al.}(2010){Gandolfi}, {H{\'e}brard}, {Alonso},
  {Deleuil}, {Guenther}, {Fridlund}, {Endl}, {Eigm{\"u}ller}, {Csizmadia},
  {Havel}, {Aigrain}, {Auvergne}, {Baglin}, {Barge}, {Bonomo}, {Bord{\'e}},
  {Bouchy}, {Bruntt}, {Cabrera}, {Carpano}, {Carone}, {Cochran}, {Deeg},
  {Dvorak}, {Eisl{\"o}ffel}, {Erikson}, {Ferraz-Mello}, {Gazzano}, {Gibson},
  {Gillon}, {Gondoin}, {Guillot}, {Hartmann}, {Hatzes}, {Jorda}, {Kabath},
  {L{\'e}ger}, {Llebaria}, {Lammer}, {MacQueen}, {Mayor}, {Mazeh}, {Moutou},
  {Ollivier}, {P{\"a}tzold}, {Pepe}, {Queloz}, {Rauer}, {Rouan}, {Samuel},
  {Schneider}, {Stecklum}, {Tingley}, {Udry}, \& {Wuchterl}}]{GH2010}
{Gandolfi}, D., {H{\'e}brard}, G., {Alonso}, R., {et~al.} 2010, \aap, 524, A55

\bibitem[{{Ge} {et~al.}(2006){Ge}, {van Eyken}, {Mahadevan}, {DeWitt}, {Kane},
  {Cohen}, {Vand en Heuvel}, {Fleming}, {Guo}, \& {Henry}}]{GE2006}
{Ge}, J., {van Eyken}, J., {Mahadevan}, S., {et~al.} 2006, \apj, 648, 683

\bibitem[{{Gilliland}(1986)}]{G1986}
{Gilliland}, R.~L. 1986, \apj, 300, 339

\bibitem[{{Goldreich} \& {Nicholson}(1989)}]{GN1989}
{Goldreich}, P. \& {Nicholson}, P.~D. 1989, \apj, 342, 1079

\bibitem[{{Goodman} \& {Dickson}(1998)}]{GD1998}
{Goodman}, J. \& {Dickson}, E.~S. 1998, \apj, 507, 938

\bibitem[{{Gough} \& {Thompson}(1990)}]{GT1990}
{Gough}, D.~O. \& {Thompson}, M.~J. 1990, \mnras, 242, 25

\bibitem[{{Granata} {et~al.}(2014){Granata}, {Nascimbeni}, {Piotto}, {Bedin},
  {Borsato}, {Cunial}, {Damasso}, \& {Malavolta }}]{GN2014}
{Granata}, V., {Nascimbeni}, V., {Piotto}, G., {et~al.} 2014, Astronomische
  Nachrichten, 335, 797

\bibitem[{{Guenel} {et~al.}(2016{\natexlab{a}}){Guenel}, {Baruteau}, {Mathis},
  \& {Rieutord}}]{G2016a}
{Guenel}, M., {Baruteau}, C., {Mathis}, S., \& {Rieutord}, M.
  2016{\natexlab{a}}, \aap, 589, A22

\bibitem[{{Guenel} {et~al.}(2016{\natexlab{b}}){Guenel}, {Mathis}, {Baruteau},
  \& {Rieutord}}]{G2016b}
{Guenel}, M., {Mathis}, S., {Baruteau}, C., \& {Rieutord}, M.
  2016{\natexlab{b}}, ArXiv e-prints, arXiv:1612.05071

\bibitem[{{H{\'e}brard} {et~al.}(2013){H{\'e}brard}, {Collier Cameron},
  {Brown}, {D{\'{\i}}az}, {Faedi}, {Smalley}, {Anderson}, {Armstrong},
  {Barros}, {Bento}, {Bouchy}, {Doyle}, {Enoch}, {G{\'o}mez Maqueo Chew},
  {H{\'e}brard}, {Hellier}, {Lendl}, {Lister}, {Maxted}, {McCormac}, {Moutou},
  {Pollacco}, {Queloz}, {Santerne}, {Skillen}, {Southworth}, {Tregloan-Reed},
  {Triaud}, {Udry}, {Vanhuysse}, {Watson}, {West}, \& {Wheatley}}]{HC2013}
{H{\'e}brard}, G., {Collier Cameron}, A., {Brown}, D.~J.~A., {et~al.} 2013,
  \aap, 549, A134

\bibitem[{{Hirano} {et~al.}(2012){Hirano}, {Sanchis-Ojeda}, {Takeda}, {Narita},
  {Winn}, {Taruya}, \& {Suto}}]{HS2012}
{Hirano}, T., {Sanchis-Ojeda}, R., {Takeda}, Y., {et~al.} 2012, \apj, 756, 66

\bibitem[{{Hut}(1980)}]{H1980}
{Hut}, P. 1980, \aap, 92, 167

\bibitem[{{Janson} {et~al.}(2008){Janson}, {Reffert}, {Brandner}, {Henning},
  {Lenzen}, \& {Hippler}}]{JR2008}
{Janson}, M., {Reffert}, S., {Brandner}, W., {et~al.} 2008, \aap, 488, 771

\bibitem[{{Jardine} \& {Unruh}(1999)}]{JU1999}
{Jardine}, M. \& {Unruh}, Y.~C. 1999, \aap, 346, 883

\bibitem[{Jur\ifmmode \check{c}\else \v{c}\fi{}i\ifmmode~\check{s}\else
  \v{s}\fi{}inov\'a {et~al.}(2013)Jur\ifmmode \check{c}\else
  \v{c}\fi{}i\ifmmode~\check{s}\else \v{s}\fi{}inov\'a, Jur\ifmmode
  \check{c}\else \v{c}\fi{}i\ifmmode~\check{s}\else \v{s}\fi{}in, Remeck\'y, \&
  Zalom}]{JJ2013}
Jur\ifmmode \check{c}\else \v{c}\fi{}i\ifmmode~\check{s}\else
  \v{s}\fi{}inov\'a, E., Jur\ifmmode \check{c}\else
  \v{c}\fi{}i\ifmmode~\check{s}\else \v{s}\fi{}in, M., Remeck\'y, R., \& Zalom,
  P. 2013, Phys. Rev. E, 87, 043010

\bibitem[{{K{\"a}pyl{\"a}} {et~al.}(2019){K{\"a}pyl{\"a}}, {Rheinhardt},
  {Brandenburg}, \& {K{\"a}pyl{\"a}}}]{KR2019}
{K{\"a}pyl{\"a}}, P.~J., {Rheinhardt}, M., {Brandenburg}, A., \&
  {K{\"a}pyl{\"a}}, M.~J. 2019, arXiv e-prints [\eprint[arXiv]{1901.00787}]

\bibitem[{{Kochukhov} {et~al.}(2017){Kochukhov}, {Petit}, {Strassmeier},
  {Carroll}, {Fares}, {Folsom}, {Jeffers}, {Korhonen}, {Monnier}, {Morin},
  {Ros{\'e}n}, {Roettenbacher}, \& {Shulyak}}]{KP2017}
{Kochukhov}, O., {Petit}, P., {Strassmeier}, K.~G., {et~al.} 2017,
  Astronomische Nachrichten, 338, 428

\bibitem[{{Lagarde} {et~al.}(2012){Lagarde}, {Decressin}, {Charbonnel},
  {Eggenberger}, {Ekstr{\"o}m}, \& {Palacios}}]{LD2012}
{Lagarde}, N., {Decressin}, T., {Charbonnel}, C., {et~al.} 2012, \aap, 543,
  A108

\bibitem[{{Landin} {et~al.}(2010){Landin}, {Mendes}, \& {Vaz}}]{LM2010}
{Landin}, N.~R., {Mendes}, L.~T.~S., \& {Vaz}, L.~P.~R. 2010, \aap, 510, A46

\bibitem[{{Lanza} {et~al.}(2011){Lanza}, {Damiani}, \& {Gandolfi}}]{LD2011}
{Lanza}, A.~F., {Damiani}, C., \& {Gandolfi}, D. 2011, \aap, 529, A50

\bibitem[{{Lehnert}(1954)}]{L1954}
{Lehnert}, B. 1954, \apj, 119, 647

\bibitem[{{Lin} \& {Ogilvie}(2018)}]{LO2018}
{Lin}, Y. \& {Ogilvie}, G.~I. 2018, \mnras, 474, 1644

\bibitem[{{Line} {et~al.}(2015){Line}, {Teske}, {Burningham}, {Fortney}, \&
  {Marley}}]{LT2015}
{Line}, M.~R., {Teske}, J., {Burningham}, B., {Fortney}, J.~J., \& {Marley},
  M.~S. 2015, \apj, 807, 183

\bibitem[{{Mancini} {et~al.}(2015){Mancini}, {Esposito}, {Covino}, {Raia},
  {Southworth}, {Tregloan-Reed}, {Biazzo}, {Bonomo}, {Desidera}, {Lanza},
  {Maciejewski}, {Poretti}, {Sozzetti}, {Borsa}, {Bruni}, {Ciceri}, {Claudi},
  {Cosentino}, {Gratton}, {Martinez Fiorenzano}, {Lodato}, {Lorenzi},
  {Marzari}, {Murabito}, {Affer}, {Bignamini}, {Bedin}, {Boccato}, {Damasso},
  {Henning}, {Maggio}, {Micela}, {Molinari}, {Pagano}, {Piotto}, {Rainer},
  {Scandariato}, {Smareglia}, \& {Zanmar Sanchez}}]{ME2015}
{Mancini}, L., {Esposito}, M., {Covino}, E., {et~al.} 2015, \aap, 579, A136

\bibitem[{{Mancini} {et~al.}(2018){Mancini}, {Esposito}, {Covino},
  {Southworth}, {Biazzo}, {Bruni}, {Ciceri}, {Evans}, {Lanza}, {Poretti},
  {Sarkis}, {Smith}, {Brogi}, {Affer}, {Benatti}, {Bignamini}, {Boccato},
  {Bonomo}, {Borsa}, {Carleo}, {Claudi}, {Cosentino}, {Damasso}, {Desidera},
  {Giacobbe}, {Gonz{\'a}lez-{\'A}lvarez}, {Gratton}, {Harutyunyan}, {Leto},
  {Maggio}, {Malavolta}, {Maldonado}, {Martinez-Fiorenzano}, {Masiero},
  {Micela}, {Molinari}, {Nascimbeni}, {Pagano}, {Pedani}, {Piotto}, {Rainer},
  {Scandariato}, {Smareglia}, {Sozzetti}, {Andreuzzi}, \& {Henning}}]{ME2018}
{Mancini}, L., {Esposito}, M., {Covino}, E., {et~al.} 2018, \aap, 613, A41

\bibitem[{{Mathis}(2015)}]{M2015}
{Mathis}, S. 2015, \aap, 580, L3

\bibitem[{{Mathis} {et~al.}(2016){Mathis}, {Auclair-Desrotour}, {Guenel},
  {Gallet}, \& {Le Poncin-Lafitte}}]{MA2016}
{Mathis}, S., {Auclair-Desrotour}, P., {Guenel}, M., {Gallet}, F., \& {Le
  Poncin-Lafitte}, C. 2016, \aap, 592, A33

\bibitem[{{Mathis} \& {Le Poncin-Lafitte}(2009)}]{ML2009}
{Mathis}, S. \& {Le Poncin-Lafitte}, C. 2009, \aap, 497, 889

\bibitem[{{Matt} {et~al.}(2015){Matt}, {Brun}, {Baraffe}, {Bouvier}, \&
  {Chabrier}}]{MB2015}
{Matt}, S.~P., {Brun}, A.~S., {Baraffe}, I., {Bouvier}, J., \& {Chabrier}, G.
  2015, \apj, 799, L23

\bibitem[{{Maxted} {et~al.}(2015){Maxted}, {Serenelli}, \&
  {Southworth}}]{MS2015}
{Maxted}, P.~F.~L., {Serenelli}, A.~M., \& {Southworth}, J. 2015, \aap, 577,
  A90

\bibitem[{{Mayor} {et~al.}(1997){Mayor}, {Queloz}, {Udry}, \&
  {Halbwachs}}]{MQ1997}
{Mayor}, M., {Queloz}, D., {Udry}, S., \& {Halbwachs}, J.-L. 1997, in IAU
  Colloq. 161: Astronomical and Biochemical Origins and the Search for Life in
  the Universe, ed. C.~{Batalli Cosmovici}, S.~{Bowyer}, \& D.~{Werthimer}, 313

\bibitem[{{Mohler-Fischer} {et~al.}(2013){Mohler-Fischer}, {Mancini},
  {Hartman}, {Bakos}, {Penev}, {Bayliss}, {Jord{\'a}n}, {Csubry}, {Zhou},
  {Rabus}, {Nikolov}, {Brahm}, {Espinoza}, {Buchhave}, {B{\'e}ky}, {Suc},
  {Cs{\'a}k}, {Henning}, {Wright}, {Tinney}, {Addison}, {Schmidt}, {Noyes},
  {Papp}, {L{\'a}z{\'a}r}, {S{\'a}ri}, \& {Conroy}}]{MMH2013}
{Mohler-Fischer}, M., {Mancini}, L., {Hartman}, J.~D., {et~al.} 2013, \aap,
  558, A55

\bibitem[{{Ogilvie}(2005)}]{O2005}
{Ogilvie}, G.~I. 2005, Journal of Fluid Mechanics, 543, 19

\bibitem[{{Ogilvie}(2009)}]{O2009}
{Ogilvie}, G.~I. 2009, \mnras, 396, 794

\bibitem[{{Ogilvie}(2013)}]{O2013}
{Ogilvie}, G.~I. 2013, \mnras, 429, 613

\bibitem[{{Ogilvie}(2014)}]{O2014}
{Ogilvie}, G.~I. 2014, \araa, 52, 171

\bibitem[{{Ogilvie} \& {Lesur}(2012)}]{OL2012}
{Ogilvie}, G.~I. \& {Lesur}, G. 2012, \mnras, 422, 1975

\bibitem[{{Ogilvie} \& {Lin}(2004)}]{OL2004}
{Ogilvie}, G.~I. \& {Lin}, D.~N.~C. 2004, \apj, 610, 477

\bibitem[{{Ogilvie} \& {Lin}(2007)}]{OL2007}
{Ogilvie}, G.~I. \& {Lin}, D.~N.~C. 2007, \apj, 661, 1180

\bibitem[{Perryman(2018)}]{P2018}
Perryman, M. 2018, The Exoplanet Handbook

\bibitem[{{Rebull} {et~al.}(2004){Rebull}, {Wolff}, \& {Strom}}]{RW2004}
{Rebull}, L.~M., {Wolff}, S.~C., \& {Strom}, S.~E. 2004, \aj, 127, 1029

\bibitem[{{Reiners}(2012)}]{R2012}
{Reiners}, A. 2012, Living Reviews in Solar Physics, 9, 1

\bibitem[{{Remus} {et~al.}(2012){Remus}, {Mathis}, \& {Zahn}}]{RM2012}
{Remus}, F., {Mathis}, S., \& {Zahn}, J.~P. 2012, \aap, 544, A132

\bibitem[{{Schlichting}(2014)}]{S2014}
{Schlichting}, H.~E. 2014, \apj, 795, L15

\bibitem[{{Schneider} {et~al.}(2011){Schneider}, {Dedieu}, {Le Sidaner},
  {Savalle}, \& {Zolotukhin}}]{SD2011}
{Schneider}, J., {Dedieu}, C., {Le Sidaner}, P., {Savalle}, R., \&
  {Zolotukhin}, I. 2011, \aap, 532, A79

\bibitem[{{See} {et~al.}(2017){See}, {Jardine}, {Vidotto}, {Donati}, {Boro
  Saikia}, {Fares}, {Folsom}, {H{\'e}brard}, {Jeffers}, {Marsden}, {Morin},
  {Petit}, {Waite}, \& {BCool Collaboration}}]{SJ2017}
{See}, V., {Jardine}, M., {Vidotto}, A.~A., {et~al.} 2017, \mnras, 466, 1542

\bibitem[{{See} {et~al.}(2015){See}, {Jardine}, {Vidotto}, {Donati}, {Folsom},
  {Boro Saikia}, {Bouvier}, {Fares}, {Gregory}, {Hussain}, {Jeffers},
  {Marsden}, {Morin}, {Moutou}, {do Nascimento}, {Petit}, {Ros{\'e}n}, \&
  {Waite}}]{SJ2015}
{See}, V., {Jardine}, M., {Vidotto}, A.~A., {et~al.} 2015, \mnras, 453, 4301

\bibitem[{{See} {et~al.}(2019){See}, {Matt}, {Folsom}, {Boro Saikia}, {Donati},
  {Fares}, {Finley}, {H{\'e}brard}, {Jardine}, \& {Jeffers}}]{SM2019}
{See}, V., {Matt}, S.~P., {Folsom}, C.~P., {et~al.} 2019, \apj, 876, 118

\bibitem[{{Skumanich}(1972)}]{S1972}
{Skumanich}, A. 1972, \apj, 171, 565

\bibitem[{{Southworth}(2011)}]{S2011}
{Southworth}, J. 2011, \mnras, 417, 2166

\bibitem[{{Spiegel} \& {Zahn}(1992)}]{SZ1992}
{Spiegel}, E.~A. \& {Zahn}, J.~P. 1992, \aap, 265, 106

\bibitem[{{Strugarek} {et~al.}(2017){Strugarek}, {Beaudoin}, {Charbonneau},
  {Brun}, \& {do Nascimento}}]{SB2017}
{Strugarek}, A., {Beaudoin}, P., {Charbonneau}, P., {Brun}, A.~S., \& {do
  Nascimento}, J.~D. 2017, Science, 357, 185

\bibitem[{{Strugarek} {et~al.}(2013){Strugarek}, {Brun}, {Mathis}, \&
  {Sarazin}}]{SB2013}
{Strugarek}, A., {Brun}, A.~S., {Mathis}, S., \& {Sarazin}, Y. 2013, \apj, 764,
  189

\bibitem[{{Terquem} {et~al.}(1998){Terquem}, {Papaloizou}, {Nelson}, \&
  {Lin}}]{TP1998}
{Terquem}, C., {Papaloizou}, J.~C.~B., {Nelson}, R.~P., \& {Lin}, D.~N.~C.
  1998, \apj, 502, 788

\bibitem[{{Tregloan-Reed} {et~al.}(2015){Tregloan-Reed}, {Southworth},
  {Burgdorf}, {Novati}, {Dominik}, {Finet}, {J{\o}rgensen}, {Maier}, {Mancini},
  {Prof}, {Ricci}, {Snodgrass}, {Bozza}, {Browne}, {Dodds}, {Gerner},
  {Harps{\o}e}, {Hinse}, {Hundertmark}, {Kains}, {Kerins}, {Liebig}, {Penny},
  {Rahvar}, {Sahu}, {Scarpetta}, {Sch{\"a}fer}, {Sch{\"o}nebeck}, {Skottfelt},
  \& {Surdej}}]{TS2015}
{Tregloan-Reed}, J., {Southworth}, J., {Burgdorf}, M., {et~al.} 2015, \mnras,
  450, 1760

\bibitem[{{Tregloan-Reed} {et~al.}(2013){Tregloan-Reed}, {Southworth}, \&
  {Tappert}}]{TS2013}
{Tregloan-Reed}, J., {Southworth}, J., \& {Tappert}, C. 2013, \mnras, 428, 3671

\bibitem[{{Vidotto} {et~al.}(2014){Vidotto}, {Gregory}, {Jardine}, {Donati},
  {Petit}, {Morin}, {Folsom}, {Bouvier}, {Cameron}, {Hussain}, {Marsden},
  {Waite}, {Fares}, {Jeffers}, \& {do Nascimento}}]{VG2014}
{Vidotto}, A.~A., {Gregory}, S.~G., {Jardine}, M., {et~al.} 2014, \mnras, 441,
  2361

\bibitem[{{Vilhu}(1984)}]{V1984}
{Vilhu}, O. 1984, \aap, 133, 117

\bibitem[{{Weber} \& {Davis}(1967)}]{WD1967}
{Weber}, E.~J. \& {Davis}, Leverett, J. 1967, \apj, 148, 217

\bibitem[{{Wei}(2016)}]{W2016}
{Wei}, X. 2016, \apj, 828, 30

\bibitem[{{Wei}(2018)}]{W2018}
{Wei}, X. 2018, \apj, 854, 34

\bibitem[{{Winn} {et~al.}(2005){Winn}, {Noyes}, {Holman}, {Charbonneau},
  {Ohta}, {Taruya}, {Suto}, {Narita}, {Turner}, {Johnson}, {Marcy}, {Butler},
  \& {Vogt}}]{WN2005}
{Winn}, J.~N., {Noyes}, R.~W., {Holman}, M.~J., {et~al.} 2005, \apj, 631, 1215

\bibitem[{{Witte} \& {Savonije}(2002)}]{WS2002}
{Witte}, M.~G. \& {Savonije}, G.~J. 2002, \aap, 386, 222

\bibitem[{{Wright} \& {Drake}(2016)}]{WD2016}
{Wright}, N.~J. \& {Drake}, J.~J. 2016, \nat, 535, 526

\bibitem[{{Zahn}(1966{\natexlab{a}})}]{Z1966a}
{Zahn}, J.~P. 1966{\natexlab{a}}, Annales d'Astrophysique, 29, 313

\bibitem[{{Zahn}(1966{\natexlab{b}})}]{Z1966b}
{Zahn}, J.~P. 1966{\natexlab{b}}, Annales d'Astrophysique, 29, 489

\bibitem[{{Zahn}(1966{\natexlab{c}})}]{Z1966c}
{Zahn}, J.~P. 1966{\natexlab{c}}, Annales d'Astrophysique, 29, 565

\bibitem[{{Zahn}(1975)}]{Z1975}
{Zahn}, J.~P. 1975, \aap, 41, 329

\bibitem[{{Zahn}(1977)}]{Z1977}
{Zahn}, J.~P. 1977, \aap, 500, 121

\bibitem[{{Zahn}(1989)}]{Z1989}
{Zahn}, J.~P. 1989, \aap, 220, 112

\bibitem[{{Zahn} \& {Bouchet}(1989)}]{ZB1989}
{Zahn}, J.~P. \& {Bouchet}, L. 1989, \aap, 223, 112

\bibitem[{{Zanni} \& {Ferreira}(2013)}]{ZF2013}
{Zanni}, C. \& {Ferreira}, J. 2013, \aap, 550, A99

\bibitem[{{Zeldovich}(1983)}]{Z1983}
{Zeldovich}, Y.~B. 1983, {Magnetic fields in astrophysics}

\end{thebibliography}

\begin{appendix}
\section{Magnetic field scaling laws}
\label{AA}
Several estimates of the magnetic field of stars can be derived thanks to  energetic or force   balances in the equations for heat transport  (items \ref{eq} and \ref{bu} below) or  for momentum (items \ref{mag}, and also \ref{eq} and \ref{bu}).
Based on the MHD equations, and assuming stationarity, the Navier-Stokes equation of convective motions  is 
\begin{equation}
\begin{aligned}
\underset{(1)}{\underline{2\rho\boldsymbol\Omega\times\bu}}+\underset{(2)}{\underline{\rho(\bu\cdot\bn)\bu}}=&-\bn p+\rho\nu\Delta{\bf u}+\underset{(3)}{\underline{\frac{\boldsymbol\nabla\times{\bf B}}{\mu_0}\times\bf B}}+\rho\bf g,
\end{aligned}
\label{eqA0}
\end{equation}
where we have introduced $\bf g$ the gravitational acceleration of the star. At the base of the convective zone, we assume that the gradient and the velocity scale like $\nabla\approx\lc^{-1}$ and $||{\bf u}||\approx\uc$.
\begin{enumerate}
\item\textit{Magnetostrophy}\label{mag}\\

In this regime, the  Lorentz force balances the Coriolis force i.e.  $(3)\approx(1)$ in  Eq. (\ref{eqA0}). Thus,
\begin{equation}
\frac{B^2}{\mu_0\lc}\approx\rho 2\Omega\uc
\Rightarrow B\approx\sqrt{2\mu_0\KE\times\Ro^{-1}},
\end{equation}
where $\KE=\rho\uc^2/2$ is the kinetic energy density associated with convective motions, and $\Ro=\uc/(2\Omega\lc)$ the convective Rossby number.\\

\item\textit{Equipartition regime}\label{eq}\\

In this regime we assume that the dynamo is efficient enough such that the convective kinetic energy density $\KE$ of the fluid is equivalent to the magnetic energy density $\ME=B^2/(2\mu_0)$ i.e. $(3)\approx(2)$ in Eq. (\ref{eqA0}):
\begin{equation}
\ME\approx\KE\Rightarrow \frac{B^2}{\mu_0}\approx\rho\uc^2\Rightarrow B\approx\sqrt{2\mu_0\KE}.
\end{equation}
\\

\item\textit{Buoyancy dynamo}\label{bu}\\

Taking the curl of Eq. (\ref{eqA0}) and neglecting viscous and inertial   forces we obtain
\begin{equation}
\rho(\bf\Omega\cdot\bn)\bf v=\bn\times\left(\frac{\bn\times B}{\mu_0}\times B\right)+\boldsymbol\nabla\rho\times\bf g,
\label{eqB0}
\end{equation}
assuming that the velocity is divergence-free.
The buoyancy dynamo lies on the fact that each term in Eq. (\ref{eqB0}) has the same order of magnitude \citep{D2013,AB2019}. If the Rossby number is small enough, namely the rotation frequency of the body is larger than the convective frequency (the inverse of the convective turnover time), the convective flows become almost columnar, and aligned along the rotation axis. In the stellar evolution grids we use, this assumption is at least verified in the first half of the convective zone from the base. As a  consequence, two characteristic length scales appear, $l_\parallel$ and $l_\perp$: these lengths are parallel and perpendicular to the rotation axis, respectively. A direct result is that $l_\perp<l_\parallel$, and we assume that $\lc\simeq l_\parallel$ so that  the convective Rossby number is now $\Ro=\uc/(2\Omega l_\parallel)$. It seems consistent with the high values of the convective length scale  we find  at the base of the envelope for different stars (several tenths as  the convective zone thickness). Thus, taking the characteristic scales of each quantities in Eq. (\ref{eqB0}):
\begin{equation}
\frac{\Omega u_c}{l_\parallel}\approx\frac{g}{l_\perp}\approx\frac{B^2}{\rho\mu_0 l_\perp^2}
\label{eqB1},
\end{equation}
where $\boldsymbol\Omega\nabla\boldsymbol\approx\Omega/l_\parallel$ by definition of $l_\parallel$, otherwise $\nabla\propto l_\perp^{-1}$ because $l_\perp<l_\parallel$.
Then, we set up $p_g=g\uc$ the buoyancy power (or rate of buoyancy work) per unit of mass. 
We can also express the magnetic energy in term of buoyancy power because of the equivalence of forces in Eq. (\ref{eqB0}). Therefore, in order to respect dimensional homogeneity, the magnetic energy per unit of mass satisfies the following relationship:
\begin{equation}
\frac{B^2}{\rho\mu_0}=l_\parallel^{2/3}p_\mathrm{g}^{2/3},
\label{eqB3}
\end{equation}
where we are only using $l_\parallel$, again because $l_\parallel>l_\perp$ since we are looking for the  large-scale dynamo magnetic field.
From the equivalences in Eq. (\ref{eqB1}), we can express $g$ as a function of $\Omega$, $\uc$, $l_\parallel$ and the energy by unit of mass $B^2/(\rho\mu_0)$:
\begin{equation}
g^2=\frac{\Omega\uc}{l_\parallel}\frac{B^2}{\rho\mu_0}.
\label{eqB4}
\end{equation}
Injecting Eq. (\ref{eqB4}) in $p_g^2=(g\uc)^2$, and all of this in Eq. (\ref{eqB3}), one can find:
\begin{equation}
\left(\frac{B^2}{\rho\mu_0}\right)^2=l_\parallel\Omega\uc^3.
\label{eqB5}
\end{equation}
Considering that $\Ro=\uc/(2\Omega l_\parallel)$, Eq. (\ref{eqB5}) can be written finally in the form:
\begin{equation}
\frac{\ME}{\KE}=\Ro^{-1/2}, \text{ or }\ B=\sqrt{2\mu_0\KE\times\Ro^{-1/2}}.
\end{equation}
\end{enumerate}
\section{Limitations of the best scaling for $B_\mathrm{dip}^{obs}$}
\label{dis}
\begin{figure}[h]
\resizebox{\hsize}{!}{\includegraphics{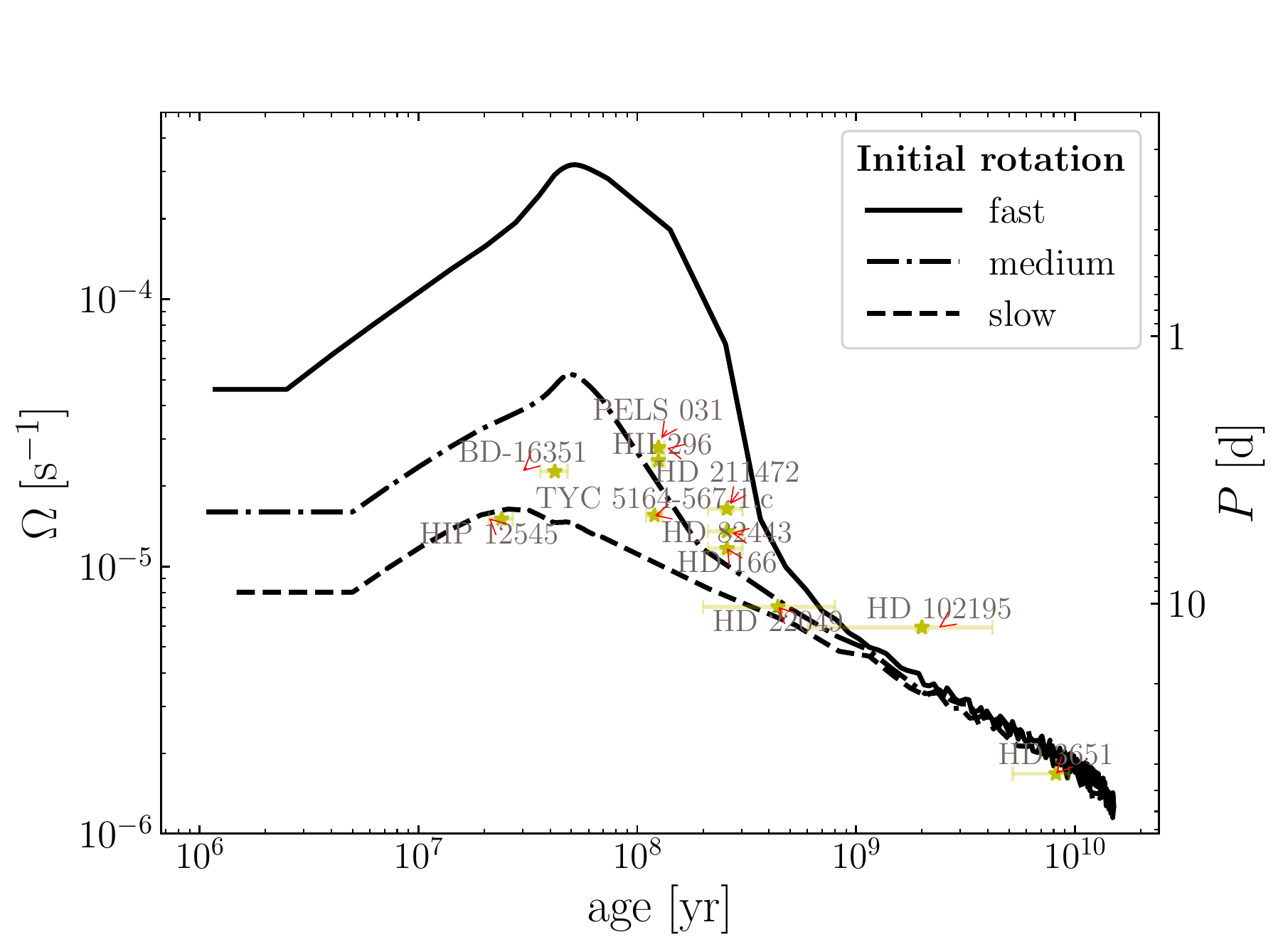}}
\caption{Rotation frequency  on the vertical left axis (and period on the vertical right axis) versus the age of  $0.9\,M_\odot$ stars ($\color{jv}\star$), given fast, median, and slow initial rotations. Ages, stellar periods and their associated error bars can be found in the papers quoted in Fig. \ref{fig1}.}
\label{Om}
\end{figure}

In Figs. \ref{fig1} and \ref{BobsBdyn}, we observe that $B_\mathrm{dip}$ calculated with the magnetostrophic regime and a fast initial rotation at the top of the convective zone is in adequacy with the observed dipolar magnetic field at the surface. We present here some limitations of this comparison. 
In Fig. \ref{Om}, we display  the angular velocity (and the period in the vertical right axis)  of the $0.9M_\odot$ stars featured in Fig. \ref{fig1}  against their age, compared to the different rotational evolutions  given by STAREVOL. The angular velocities are found in \cite{SJ2017} and references therein. They are obtained mainly through photometry while the age of the stars are mostly determined by the comparison between isochrones and colour-magnitude diagrams. Note that the error bars of the angular velocities are too small to be visible on the graph. We point out that the rotation periods of the stars seem to be in agreement with the slow or median rotation profiles provided by STAREVOL, unlike what we observed for the surface dipolar magnetic fields in Fig. \ref{fig1}. To some extent, the same assessment is made when displaying rotation versus time diagram for other low-mass stars (not presented here). There are several possible explanations for this result. The discrepancy could be first attributed to the simplicity of our dynamo-induced magnetic fields models. Then, it is possible that the metallicity affects the rotation evolution models as it has been demonstrated by Amard et al. in prep. (we remind the reader that the metallicity is fixed in Sect. \ref{sec23} to the solar value). Furthermore, as we said in Sect. \ref{sec24}, this observation tends to be more nuanced with increasing data.
As a result, we have to cautiously interpret the fact that $B_\mathrm{dip}$ estimated with the magnetostrophic regime and a fast initial rotation match the observed values of the dipolar magnetic field at the surface.

\longtab[1]{
\begin{landscape}
\begin{longtable}{c||cccccc}
\caption{Star-planet systems with a quasi-circular and quasi-coplanar orbit, $e<0.1$ and $|\lambda|<30^\circ$ with $e$ and $\lambda$ the eccentricity and the sky-projected  spin-orbit misalignment angle, respectively. All the planets are hot Jupiters, denoted (b), and characterised by an orbital period $P_\mathrm{o}$  found in the \url{https://www.exoplanet.eu} catalogue, along with the age, the radius $R$, and the mass $M_\star$ of the host star. The errors of these parameters (coming from the catalogue) generally refer to $1\sigma$ errors \citep{SD2011}. 
The references are for the  stellar rotation period $P_\star$.}\\
\label{tabsys}
Systems & $M_\star\ [M_\odot]$ & $\mathrm{age}\ [\mathrm{Gyr}]$ & $R\ [R_\odot]$ & $P_\mathrm{o}\ [\mathrm{d}]$ & $P_\star\ [\mathrm{d}]$ & References\\
\hline
\rule[0.8ex]{0pt}{2ex}
CoRoT-11 (b) & $1.27\pm0.05$ & $2.0\pm1.0$&$1.43\pm0.033$ & $2.994325\pm2.1\e{-5}$ & $1.73\pm0.22$&1 \\
\rule[0.8ex]{0pt}{2ex}
HAT-P-3 (b) & $0.917\pm0.03$ & $1.6\pm1.3$&$0.827\pm0.055$ & $2.899703\pm5.4\e{-5}$ & $20.2\pm2.0$&2 \\
\rule[0.8ex]{0pt}{2ex}
HAT-P-4 (b) & $1.26\pm0.14$ & $4.2^{+2.6}_{-0.6}$&$1.27\pm0.05$ & $3.0565114\pm2.8\e{-6}$ & $14.0^{+1.0}_{-0.9}$&3 \\
\rule[0.8ex]{0pt}{2ex}
HAT-P-8 (b) & $1.28\pm0.04$ & $3.4\pm1.0$&$1.5\pm0.06$ & $3.0763402\pm1.5\e{-6}$ & $5.9^{+0.5}_{-0.4}$&3 \\
\rule[0.8ex]{0pt}{2ex}
HAT-P-9 (b) & $1.28\pm0.13$ & $1.6\pm1.4$&$1.4\pm0.06$ & $3.922814\pm2\e{-6}$ & $5.61\pm0.78$&4 \\
\rule[0.8ex]{0pt}{2ex}
HAT-P-13 (b) & $1.22\pm0.1$ & $5.0\pm0.8$&$1.28\pm0.079$ & $2.916243\pm3\e{-6}$ & $47.4^{+14.1}_{-8.9}$&3 \\
\rule[0.8ex]{0pt}{2ex}
HAT-P-16 (b) & $1.218\pm0.039$ & $2.0\pm0.8$&$1.289\pm0.066$ & $2.77596\pm3\e{-6}$ & $16.0^{+4.1}_{-2.8}$&3 \\
\rule[0.8ex]{0pt}{2ex}
HAT-P-20 (b) & $0.756\pm0.028$ & $6.7\pm3.8$&$0.867\pm0.033$ & $2.875317\pm4\e{-6}$ & $14.48\pm0.02$&5, 6 \\
\rule[0.8ex]{0pt}{2ex}
HAT-P-22 (b) & $0.916\pm0.035$ & $12.4\pm2.6$&$1.08\pm0.058$ & $3.21222\pm9\e{-6}$ & $28.7\pm0.4$&2 \\
\rule[0.8ex]{0pt}{2ex}
HAT-P-24 (b) & $1.191\pm0.042$ & $2.8\pm0.6$&$1.242\pm0.067$ & $3.35524\pm7\e{-6}$ & $6.67\pm0.68$&4 \\
\rule[0.8ex]{0pt}{2ex}
HAT-P-27 (b) & $0.945\pm0.035$ & $4.4\pm2.6$&$1.055\pm0.036$ & $3.0395803\pm1.5\e{-6}$ & $17.8\pm7.8$&7 \\
\rule[0.8ex]{0pt}{2ex}
HAT-P-36 (b) & $1.022\pm0.049$ & $6.6\pm1.8$&$1.264\pm0.071$ & $1.327347\pm3\e{-6}$ & $15.3\pm0.4$&8 \\
\rule[0.8ex]{0pt}{2ex}
HATS-2 (b) & $0.882\pm0.037$ & $9.7\pm2.9$&$1.168\pm0.03$ & $1.354133\pm1\e{-6}$ & $24.98\pm0.04$&9, 10 \\
\rule[0.8ex]{0pt}{2ex}
HD 149026 (b) & $1.3\pm0.1$ & $2.0\pm0.8$&$0.718\pm0.065$ & $2.8758916\pm1.4\e{-6}$ & $10.0^{+1.2}_{-1.0}$&3 \\
\rule[0.8ex]{0pt}{2ex}
HD 209458 (b) & $1.148\pm0.022$ & $4.0\pm2.0$&$1.38\pm0.018$ & $3.52472\pm2.82\e{-5}$ & $14.4\pm2.1$&20, 21 \\
\rule[0.8ex]{0pt}{2ex}
KELT-1 (b) & $1.335\pm0.063$ & $1.75\pm0.25$&$1.15^{+0.1}_{-0.16}$ & $1.217514\pm1.5\e{-5}$ & $1.35\pm0.04$&11 \\
\rule[0.8ex]{0pt}{2ex}
Kepler-8 (b) & $1.213\pm0.063$ & $3.84\pm1.5$&$1.419\pm0.058$ & $3.52254\pm5\e{-5}$ & $7.5\pm0.3$&12 \\
\rule[0.8ex]{0pt}{2ex}
TrES-2 (b) & $0.98\pm0.062$ & $5.1\pm2.7$&$1.189\pm0.025$ & $2.4706133738\pm1.87\e{-8}$ & $48.7^{+56.0}_{-17.8}$&3 \\
\rule[0.8ex]{0pt}{2ex}
TrES-4 (b) & $1.388\pm0.042$ & $2.9\pm0.3$&$1.706\pm0.056$ & $3.5539268\pm3.2\e{-6}$ & $10.7^{+1.7}_{-1.3}$&3 \\
\rule[0.8ex]{0pt}{2ex}
WASP-5 (b) & $1.0\pm0.06$ & $3.0\pm1.4$&$1.171\pm0.057$ & $1.6284246\pm1.3\e{-6}$ & $16.2\pm0.4$&9 \\
\rule[0.8ex]{0pt}{2ex}
WASP-6 (b) & $0.888\pm0.08$ & $11.0\pm7.0$&$1.224^{+0.051}_{-0.052}$ & $3.361006\pm3.5\e{-6}$ & $23.8\pm0.15$&15 \\
\rule[0.8ex]{0pt}{2ex}
WASP-14 (b) & $1.211\pm0.122$ & $0.75\pm0.25$&$1.281^{+0.075}_{-0.082}$ & $2.2437661\pm1.1\e{-6}$ & $23.68\pm6.35$&4 \\
\rule[0.8ex]{0pt}{2ex}
WASP-16 (b) & $1.022\pm0.101$ & $2.3\pm2.2$&$1.008\pm0.071$ & $3.1186009\pm1.31\e{-5}$ & $33.8^{+8.9}_{-6.1}$&3 \\
\rule[0.8ex]{0pt}{2ex}
WASP-18 (b) & $1.24\pm0.04$ & $0.63\pm0.53$&$1.165\pm0.077$ & $0.9414518\pm4\e{-7}$ & $5.6\pm1.11$&4 \\
\rule[0.8ex]{0pt}{2ex}
WASP-19 (b) & $0.904\pm0.045$ & $11.5\pm2.7$&$1.395\pm0.025$ & $0.78884\pm3\e{-7}$ & $11.76\pm0.09$&9, 16 \\
\rule[0.8ex]{0pt}{2ex}
WASP-20 (b) & $1.202\pm0.04$ & $7.0^{+1.0}_{-2.0}$&$1.459\pm0.057$ & $4.8996285\pm3.4\e{-6}$ & $8.1^{+3.0}_{-2.7}$&3 \\
\rule[0.8ex]{0pt}{2ex}
WASP-28 (b) & $1.021\pm0.05$ & $5.0^{+2.0}_{-3.0}$&$1.213\pm0.042$ & $3.40883\pm6\e{-6}$ & $13.0^{+2.3}_{-1.8}$&3 \\
\rule[0.8ex]{0pt}{2ex}
WASP-52 (b) & $0.87\pm0.03$ & $0.4\pm0.3$&$1.27\pm0.03$ & $1.7497798\pm1.2\e{-6}$ & $11.8\pm3.3$&18 \\
\rule[0.8ex]{0pt}{2ex}
WASP-111 (b) & $1.5\pm0.11$ & $2.6\pm0.6$&$1.442\pm0.094$ & $2.310965\pm3.4\e{-6}$ & $8.4\pm0.7$&19 \\
\label{tab:sys}

\end{longtable}
\tablebib{(1) \citet{GH2010}; (2) \citet{ME2018}; (3) \citet{B2014}; (4) \citet{LD2011}; (5) \citet{EC2017}; (6) \citet{GN2014}; (7) \citet{AB2011}; (8) \citet{ME2015}; (9) \citet{MS2015}; (10) \citet{MMH2013}; (11) \citet{FT2015}; (12) \citet{HS2012}; (15) \citet{TS2015}; (16) \citet{TS2013}; (18) \citet{HC2013}; (19) \citet{AB2014}; (20) \citet{WN2005}; (21) \citet{CS2017}.}\end{landscape}}
\end{appendix}
\end{document}